\documentclass[
    reprint,
    amsmath,
    amssymb,
    aps,
    prd,
    superscriptaddress,
    floatfix,
    nofootinbib, 
    longbibliography
]{revtex4-2}

\usepackage{graphicx}
\usepackage{dcolumn}
\usepackage{bm}
\usepackage[utf8]{inputenc}
\usepackage[T1]{fontenc}
\usepackage{mathtools}

\usepackage{physics}

\usepackage{mathrsfs}
\usepackage{cancel}
\usepackage{siunitx}

\allowdisplaybreaks

\usepackage{booktabs}
\usepackage{tabularx}
\usepackage{comment}

\usepackage{microtype}

\usepackage{xcolor}
\definecolor{apslink}{RGB}{0,84,166}

\usepackage{hyperref}

\hypersetup{
    colorlinks   = true,
    linkcolor    = apslink,
    citecolor    = apslink,
    urlcolor     = apslink,
    breaklinks   = true,
    pdfstartview = {FitH}
}

\usepackage[capitalize]{cleveref}
\AtBeginDocument{\RenewCommandCopy\qty\SI}

\begin{document}

\title{Constraints on ultralight dark matter couplings from orbital perturbations of binary pulsars}

\author{Federico Huxhagen}
\email{fe.huxhagen@df.uba.ar}
\affiliation{Universidad de Buenos Aires, Facultad de Ciencias Exactas y Naturales, Departamento de Física, Buenos Aires, Argentina}
\affiliation{CONICET, Instituto de Física de Buenos Aires (IFIBA), Buenos Aires, Argentina}
\affiliation{LTE, Observatoire de Paris, Université PSL, CNRS, Sorbonne Université, Université de Lille, LNE, 61 Avenue de l’Observatoire, 75014 Paris, France}

\author{Diana L\'opez Nacir}
\email{dnacir@df.uba.ar}
\affiliation{Universidad de Buenos Aires, Facultad de Ciencias Exactas y Naturales, Departamento de Física, Buenos Aires, Argentina}
\affiliation{CONICET, Instituto de Física de Buenos Aires (IFIBA), Buenos Aires, Argentina}

\date{\today}

\begin{abstract}
Ultralight dark matter coupled directly to ordinary matter perturbs the orbits of binary pulsars, and the perturbation grows secularly when its frequency matches a harmonic of the orbital motion. We use this effect to constrain three couplings with the wideband timing data for the $50$ binary pulsars of the NANOGrav 15 yr data set: a linear and a quadratic universal coupling of a scalar field to matter, and a vector field coupled to baryon minus lepton number ($B-L$). For every system and for masses between $10^{-23}$ and $10^{-18}\,$eV, we build the timing signature from the perturbed orbital elements. We then sample the coupling together with the local amplitude, phase and direction of the field and the red noise of the pulsar, marginalizing analytically over the timing model, and combine the single-pulsar likelihoods into $95\%$ credible upper limits at each mass. For a universal linear coupling, the orbital resonances of the eccentric systems bring the limit close to the Cassini bound. For the quadratic coupling, treating the neutron star as a point particle, the limits are the strongest available between $2\times10^{-22}$ and $2\times10^{-21}\,$eV. In the case of vector ultralight dark matter, the limit is comparable to the PPTA and E\"ot-Wash bounds between $3\times10^{-22}$ and $10^{-21}\,$eV and reaches or beats the E\"ot-Wash bound at isolated orbital resonances, while MICROSCOPE remains stronger over the entire mass range.
\end{abstract}

\maketitle

\section{Introduction}
\label{sec:intro}

Dark matter accounts for about $27\%$ of the energy content of the Universe and drives the formation of cosmic structure, yet its microscopic nature remains unknown \cite{Hui2017, newfrontiers_lisanti}. One well-motivated possibility is that it is made of bosons much lighter than an electronvolt, known as ultralight dark matter (ULDM) \cite{Hu2000, Ferreira2021}. Light scalars and vectors of this kind arise in many extensions of the Standard Model, and several non-thermal mechanisms can produce them with the observed abundance \cite{Marsh2016, Antypas2022}. Because the occupation numbers in the Galactic halo are enormous, ULDM behaves as a classical field that oscillates at a frequency set by its mass $m$. This field is a superposition of many waves with random phases, so the local dark matter density sets its mean square amplitude, while the amplitude and phase at a given place are random \cite{Foster2018}.

Gravity is the only interaction of dark matter established so far, but ULDM may also couple directly to ordinary matter, and we consider three ways in which it can. The first is a scalar field coupled linearly and universally to matter, so that the masses of all bodies depend on the local value of the field. This dilaton-like interaction arises in scalar--tensor theories of gravity and is equivalent at leading order to a conformal coupling of the field to the trace of the matter stress-energy tensor \cite{T_Damour_1992}. Being universal, it respects the weak equivalence principle and is probed by tests of general relativity in the Solar System \cite{Will2014}. The second is a quadratic coupling of the same scalar. If the interaction is invariant under $\Phi \to -\Phi$, odd powers of the field are forbidden and $\Phi^2$ becomes the leading term \cite{OlivePospelov2008}. The perturbation then oscillates at twice the field frequency, and the coupling also modifies the field profile around dense bodies \cite{Hees2018}.

The third possibility is a vector field, the massive gauge boson of $U(1)_{B-L}$ \cite{Davidson1979, MohapatraMarshak1980}. Its coupling depends on the baryon and lepton content of each body, so it violates the equivalence principle. As dark matter, it exerts an oscillating, composition-dependent force on ordinary matter, which can be probed with torsion balances and atom interferometers \cite{delta_c}, satellite tests of the equivalence principle \cite{Touboul2022, Amaral:2024vwd} as well as pulsar timing \cite{Hu2026}.

Millisecond pulsars are among the most stable clocks known. Timing their pulses has tested general relativity in the strong-field regime \cite{Kramer2021, FreireWex2024}, and pulsar timing arrays (PTAs) have found evidence for a nanohertz gravitational-wave background \cite{nano_GW, EPTA2023, Reardon2023, Xu2023}. ULDM can leave several imprints on the same data. Through gravity alone, the oscillating pressure of the field makes the metric oscillate at frequency $2m$ \cite{KhmelnitskyRubakov2014}; PTAs have searched for this signal \cite{Porayko:2018sfa, Smarra:2023rjx, Afzal2023}, and it has been extended to vector and tensor fields \cite{Unal2022}. A direct coupling adds further signals: the field modulates the spin of each pulsar and the rate of the terrestrial clocks, and a $B-L$ vector imparts an acceleration on the Earth and on the pulsars \cite{smarra2024constraints, Xue:2021dp}. 

Binary pulsars give access to a further channel: the orbit. A field coupled to matter perturbs the relative motion of the two stars, and the perturbation accumulates in the orbital elements that the timing model measures. When the field frequency is close to a multiple of the orbital frequency, the perturbation is resonantly amplified and the orbital elements drift secularly. Orbital periods between hours and months place these resonances between $10^{-22}$ and $10^{-18}\,$eV. The secular regime has been studied for the linear and quadratic scalar couplings \cite{Blas2017Resonates, Blas2019}, for $B-L$ vectors \cite{spin1} and for spin-2 ULDM \cite{Armaleo:2019txp}. In these works, bounds were derived at isolated resonant masses by taking the reported uncertainty of a secular orbital parameter of individual systems, such as the derivative of the orbital period, as an upper limit on the dark matter contribution.

Subsequent studies have gone beyond the secular approximation. Reference \cite{Kus2024} developed a Bayesian method to estimate the sensitivity of binary pulsars to a linearly coupled scalar over a continuous range of masses, combining many systems, and Ref. \cite{Huxhagen2026} extended it to the quadratic scalar and to the $B-L$ vector, covering the three couplings studied here. Reference \cite{FosterBlas2025} developed a fully time-resolved description of binary resonances to forecast the reach of binaries to microhertz gravitational waves and to ULDM, and Ref. \cite{DeepNeuralULDM} explored a neural-network approach. All of these are forecasts. References \cite{Kus2024, Huxhagen2026} were built from tabulated timing precisions rather than from the arrival times, assumed white noise with a uniform cadence and used a reduced timing model. They estimated the sensitivity from one orbital parameter at a time, treated circular and eccentric systems separately, and characterized the reach for a representative configuration of the stochastic field. Red noise, the full timing model with its dispersion-measure and secular parameters, and the unknown local amplitude, phase and direction of the field are all relevant in a real measurement. No previous estimate incorporated all of these ingredients in a single analysis.

In this work, we analyze the arrival times of the $50$ binary pulsars of the NANOGrav 15 yr data set \cite{NANOGrav2023, NANOGrav15data} with a model that includes the orbital signature of the three couplings. We model the red noise of each pulsar and marginalize over its timing model and over the local realization of the field. The signal includes the perturbation of all the orbital parameters at once, and circular and eccentric systems enter the same combined limit. Here, we place upper bounds on the three couplings between $10^{-23}$ and $10^{-18}\,$eV. To our knowledge, these are the first constraints on ULDM obtained from the arrival times themselves in the orbital channel over a continuous range of masses, and they extend pulsar-timing constraints above the masses covered by existing PTA analyses.

The paper is organized as follows. In Sec. \ref{sec:theory} we describe how ULDM enters binary pulsar timing: Sec. \ref{sec:pulsars} reviews the timing models and defines the signal as a perturbation of the fitted orbital parameters, Sec. \ref{sec:couplings} introduces the field and the three couplings, and Sec. \ref{sec:orbital} describes the resulting orbital perturbations. Section \ref{sec:data} presents the data and the analysis: the NANOGrav 15 yr data set and our sample of binaries in Sec. \ref{sec:dataset}, the noise and signal model in Sec. \ref{sec:noise}, and the Bayesian method that combines the systems in Sec. \ref{sec:bayes}. We present the upper limits for the three couplings in Sec. \ref{sec:results} and conclude in Sec. \ref{sec:conclusions}. Appendix \ref{app:osculating} collects the osculating equations and the Fourier expansion of the Keplerian orbit, Appendix \ref{app_orbital} gives the perturbed orbital parameters for each coupling in the circular and eccentric bases, Appendix \ref{app:sample} lists the sample together with the masses and inclinations used by the vector template, Appendix \ref{app:pbdot} examines the signal near the exact resonance and the treatment of $\dot P_b$, and Appendix \ref{app:injection} presents the injection--recovery tests of the analysis. Throughout this work we use natural units, $\hbar = c = 1$.

\section{Ultralight dark matter and binary pulsar timing}
\label{sec:theory}
\subsection{Timing models}
\label{sec:pulsars}

Timing models describe the time of arrival of pulses at Earth, accounting for the many effects present between the pulse's emission and its observation, and for binary pulsars they give precise access to the orbital dynamics. A perturbation absent from the fitted model---such as a direct coupling between ULDM and ordinary matter---induces timing residuals.

Here we review the key features of the Blandford--Teukolsky (BT) model \cite{BlandfordTeukolsky1976}, the standard framework for mapping the binary orbital dynamics to pulse arrival times while accounting for the main relativistic effects. We keep only what is essential to define the signal; detailed treatments can be found in Refs. \cite{Kus2024, Huxhagen2026}. Let $T_p$ be the proper time measured at the pulsar. Its rotational phase, measured in turns, evolves as
\begin{equation}
     N = N_0 + \nu \, T_p + \dot{\nu} \frac{T_p^2}{2} + \ddot{\nu} \frac{T_p^3}{6},
     \label{eq: n-simo pulso}
\end{equation}
where $\nu$ is the rotation frequency, $N_0$ an arbitrary constant, and the $N$-th pulse is emitted when the phase reaches the integer $N$. The proper time can be expressed as a function of the infinite-frequency barycentric time of arrival $t$, the pulsar's intrinsic parameters and the orbital parameters.

\begin{figure}[h]
    \centering
    \includegraphics[width=0.9\linewidth]{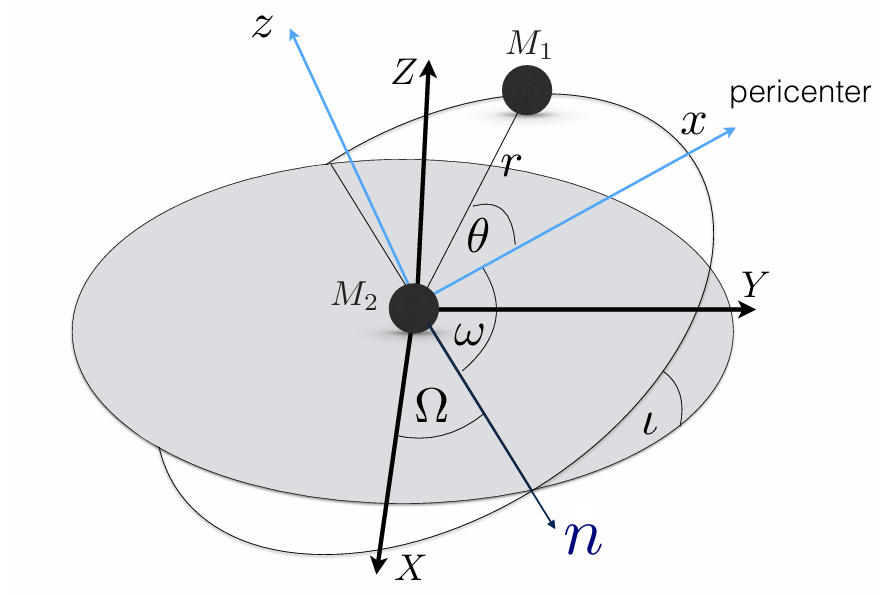}
    \caption{Description of Keplerian motion in terms of the orbital elements, defined with respect to the fundamental reference frame $(X,Y,Z)$. The associated Cartesian $(x,y,z)$ and polar $(r,\theta,z)$ coordinate systems are also indicated. $n$ denotes the line of nodes.}
    \label{fig: orbita kepleriana}
\end{figure}

Suppose that we have an initial fit of the data performed with \texttt{PINT} \cite{Luo_2021}, such as the timing solution published by NANOGrav. We denote the initial estimates of the involved parameters $N^{(1)}$, $\nu^{(1)}$, and so on. The time residuals $R(t)$, which represent deviations of the times of arrival of pulses with respect to those expected from the best-fit timing solution, are therefore given by
\begin{equation}
    - \nu^{(1)} \, R(t) = N(t, N_0, \nu,...) \, - N(t, N_0^{(1)}, \nu^{(1)},...),
\end{equation}
where the variant without superscript denotes the pulsar's true phase. Assuming that the fitted values are good estimates of the true ones, we can expand the time residuals as a function of small deviations of those parameters: 
\begin{equation}
- \nu^{(1)} R(t) = \left. \frac{\partial N}{\partial N_0} \right|_1 \delta N_0 + \left. \frac{\partial N}{\partial \nu} \right|_1 \delta \nu   +  \dots,
\label{eq: residuos expandidos}
\end{equation}
where $\delta N_0, \delta \nu, \dots$ are estimates of the parameter corrections. To first order in these corrections, $\nu^{(1)}$ and $\nu$ can be interchanged in the prefactor, and we drop the superscript in what follows.

The orbital sector of the Damour--Deruelle formalism \cite{DamourDeruelle1986}, which extends the BT model through a more complete treatment of post-Newtonian orbital dynamics, is parametrized by the projected semi-major axis $x \equiv a_1 \sin\iota$, the eccentricity $e$, the argument of periastron $\omega$, the time of periastron passage $T_0$ and the orbital period $P_b$, with the orbital geometry illustrated in Fig. \ref{fig: orbita kepleriana}. Here, $a_1 = \frac{\text{M}_2}{\text{M}_1+\text{M}_2} \, a$ is the pulsar's semi-major axis, $a$ is the semi-major axis of the relative orbit, $\iota$ is the orbital inclination, and $\text{M}_1$ and $\text{M}_2$ are the masses of the pulsar and its companion, respectively. The corresponding timing residual can then be written as
\begin{equation}
\begin{aligned}
R(t) = \, &\delta K - \frac{\partial N}{\partial x } \, \frac{\delta x }{\nu} - \frac{\partial N}{\partial e } \, \frac{\delta e }{\nu} - \frac{\partial N}{\partial \omega } \, \frac{\delta \omega }{\nu} \\ &- \frac{\partial N}{\partial T_0} \frac{\delta T_0}{\nu} - \frac{\partial N}{\partial P_b} \frac{\delta P_b}{\nu},
\end{aligned}
    \label{eq: modelo DD}
\end{equation}
where $\delta K$ collects the corrections to the parameters that do not describe the orbit, such as $\delta N_0$ and $\delta \nu$. The partial derivatives are evaluated at the best-fit values. For nearly circular orbits both $\omega$ and $T_0$ become ill defined, since the periastron itself is, and the ELL1 model \cite{Lange2001} replaces them by the Laplace parameters $\eta = e \sin\omega$ and $\kappa = e \cos\omega$, together with the time of ascending-node passage $T_{\rm asc}$,
\begin{equation}
\begin{aligned}
R(t) = &\, \delta K - \frac{\partial N}{\partial x } \, \frac{\delta x }{\nu} - \frac{\partial N}{\partial \eta } \, \frac{\delta \eta}{\nu} - \frac{\partial N}{\partial \kappa } \, \frac{\delta \kappa }{\nu} \\ &- \frac{\partial N}{\partial T_{\rm asc}} \frac{\delta T_{\rm asc}}{\nu} - \frac{\partial N}{\partial P_b} \frac{\delta P_b}{\nu}.
\end{aligned}
    \label{eq: modelo ELL1}
\end{equation}

These are the five Keplerian parameters that the ULDM field perturbs, in one basis or the other depending on the eccentricity of the system. A perturbation that is not included in the timing model therefore reaches the residuals through these same partial derivatives,
\begin{equation}
    h(t) = \sum_{\mu} \frac{\partial R}{\partial p_\mu} \, \delta p_\mu(t),
    \label{eq:template}
\end{equation}
where $\mu$ indexes the five orbital parameters of the corresponding basis and $\delta p_\mu(t)$ are the variations induced by the ultralight dark matter field. The orbital period requires one modification, because it enters the residual through the accumulated orbital phase, and we return to it in Sec. \ref{sec:orbital}.

\subsection{Ultralight dark matter and its coupling to matter}
\label{sec:couplings}

How a binary pulsar responds to the oscillating field is controlled by two scales. The first is the de Broglie wavelength,
\begin{equation}
    \lambda_{\rm dB} \sim 1.3\times10^{12}\,\text{km}
    \left(\frac{10^{-3}}{v_0}\right)
    \left(\frac{10^{-18}\,\text{eV}}{m}\right),
    \label{eq:lambda_db}
\end{equation}
with $v_0 \simeq 10^{-3}$ the typical velocity of dark matter near the Sun \cite{newfrontiers_lisanti}. The systems we analyze span a broad range of orbital separations, yet even at the upper end of our mass range, the field varies on scales orders of magnitude larger than the widest orbit. Its spatial gradient is therefore irrelevant, and each binary is driven by a field that depends on time alone. The second scale is the coherence time,
\begin{equation}
    t_{\rm coh} \sim 65\,\text{yr}
    \left(\frac{10^{-3}}{v_0}\right)^{2}
    \left(\frac{10^{-18}\,\text{eV}}{m}\right),
    \label{eq:tcoh}
\end{equation}
which comfortably exceeds the $16\,$yr baseline of the data set everywhere in our mass range.

The field at the position of a binary is the superposition of a large number of modes with uncorrelated phases, so its amplitude and phase drift over a time $t_{\rm coh}$. Over an observing campaign much shorter than that drift, however, a given system does not sample the distribution: it sees one draw from it. We therefore write the scalar field seen by a binary as
\begin{equation}
    \Phi(t) = \frac{\sqrt{2\rho_{\rm DM}}}{m}\,
              \varrho \cos\!\left(mt + \Upsilon\right),
    \label{eq:campo escalar}
\end{equation}
with $\rho_{\rm DM} \simeq 0.3\,{\rm GeV\,cm^{-3}}$ the local dark matter density \cite{deSalas2021}. The dimensionless amplitude $\varrho$ and the phase $\Upsilon$ are therefore constants of each system: $\Upsilon$ is uniform on $[0,2\pi)$ and $\varrho$ follows the Rayleigh distribution
\begin{equation}
    P(\varrho) = 2\,\varrho\,e^{-\varrho^{2}},
    \label{eq:rayleigh}
\end{equation}
which follows from that superposition \cite{Foster2018, StochasticFluctuations}. Throughout this work we adopt the uncorrelated regime, in which different binaries lie in different coherence patches and each sees an independent realization of $(\varrho,\Upsilon)$, as in Refs. \cite{Kus2024, Huxhagen2026}.

We take $\Phi$ to be a dilaton-like scalar field coupled directly and universally to ordinary matter. In the point-particle description of a binary, the leading consequence of this interaction is that the component masses depend on the local value of the scalar field. Expanding about $\Phi=0$, we consider the two lowest-order possibilities,
\begin{equation}
\mathrm{M}_i(\Phi)
= \overline{\mathrm{M}}_i\left(1+\alpha\Phi\right), \, 
\mathrm{M}_i(\Phi)
= \overline{\mathrm{M}}_i
\left(1+\frac{\beta\Phi^2}{2}\right),
\label{eq:scalar_mass_couplings}
\end{equation}
where $\overline{\mathrm{M}}_i$ are the masses in the absence of the field. We study universal couplings, with $\alpha$ or $\beta$ the same for both components of the binary, and take $\beta>0$, avoiding the non-perturbative effects---such as spontaneous scalarization---that a large negative $\beta$ can trigger in neutron stars \cite{Blas2019}.

For both scalar couplings, the perturbing acceleration on the relative coordinate derived from the Euler--Lagrange equations takes the same form,
\begin{equation}
   \vec{f} = -\zeta(\Phi)\,\dot\Phi\,\dot{\vec r},
    \qquad
    \zeta(\Phi) =
    \begin{cases}
        \alpha, & \text{linear},\\[2pt]
        \beta\,\Phi, & \text{quadratic},
    \end{cases}
    \label{eq:fuerza escalar}
\end{equation}
together with the corresponding modulation of $\mathrm{M}_{\rm T}$, the total mass of the binary. The force is directed along the relative velocity, so it acts as an alternating drag whose strength does not depend on the composition of the bodies.

Spin-1 dark matter couples to matter in a qualitatively different way. We take the field to be the massive gauge boson of $U(1)_{B-L}$, which, unlike a vector coupled to the electric charge, exerts a force on electrically neutral bodies. Over time periods shorter than the coherence time, we model the field as oscillating along a fixed direction $\hat n$,
\begin{equation}
    \vec{A}(t) = -\frac{\sqrt{2\rho_{\rm DM}}}{m}\,
                 \varrho \cos\!\left(mt+\Upsilon\right)\hat n,
    \label{eq:campo vectorial}
\end{equation}
where $\varrho$ and $\Upsilon$ are distributed as in the scalar case and $\hat n$ is drawn isotropically. We give its components in the Cartesian frame $(x,y,z)$ of Fig. \ref{fig: orbita kepleriana}, which is attached to the orbit ($\hat x$ points to the periastron and $\hat z$ along the orbital angular momentum), so that
\begin{equation}
    \hat n = \sin\vartheta\cos\varphi\,\hat x+\sin\vartheta\sin\varphi\,\hat y+\cos\vartheta\,\hat z .
    \label{eq:nhat}
\end{equation}

Each body carries an effective charge $q_i$, and the interaction $\mathcal{L}_{\rm int} = \sum_i q_i\,\vec v_i \cdot \vec A$ yields a perturbing acceleration on the relative coordinate
\begin{equation}
    \vec{f} = -\frac{g\,\Delta c}{m_n}\sqrt{2\rho_{\rm DM}}\;\varrho \sin\!\left(mt+\Upsilon\right)\hat n,
    \label{eq:fuerza vectorial}
\end{equation}
with $g$ the gauge coupling, $m_n$ the neutron mass and $c_i = q_i m_n / (g\,\mathrm{M}_i)$ the charge-to-mass ratio of each component in units of the neutron, which depends on its internal composition and binding energy \cite{Nordtvedt1968, spin1}. Only the difference $\Delta c\equiv c_1-c_2$ affects the relative orbit, because any acceleration common to both bodies cancels from their relative motion. The binary therefore acts as a fifth-force experiment in which the neutron star and its companion serve as the two test bodies. In an electrically neutral body each electron cancels the $B-L$ charge of a proton, so $c_i$ is essentially the fraction of the mass carried by neutrons: $c_1\simeq1$ for the neutron star, $c_2\simeq0.5$ for a white dwarf made of helium, carbon or oxygen, whose nuclei have as many neutrons as protons, and $c_2<0.5$ for a hydrogen-rich companion.

The osculating equations of Appendix \ref{app:osculating} take the force in the polar basis $(\hat r,\hat\theta,\hat z)$ of Fig. \ref{fig: orbita kepleriana}, which moves with the pulsar. Projecting the force per unit mass of Eq. \eqref{eq:fuerza vectorial} on this basis gives
\begin{equation}
    \vec f = -F_0\sin(mt+\Upsilon)
    \begin{pmatrix}
        \sin\vartheta\cos(\varphi-\theta) \\
        \sin\vartheta\sin(\varphi-\theta) \\
        \cos\vartheta
    \end{pmatrix},
    \label{eq:fuerza vectorial polar}
\end{equation}
with $F_0= \frac{g\,\Delta c}{m_n}\sqrt{2\rho_{\rm DM}}\;\varrho$.

The scalar perturbation of Eq. \eqref{eq:fuerza escalar} is tied to the orbital motion through $\dot{\vec r}$ and stays in the orbital plane. The vector force follows the field instead, so its normal component $F_z$ perturbs the inclination and the longitude of the ascending node.

We model only the orbital perturbation. We neglect the delays that the couplings induce through the pulsar spin \cite{smarra2024constraints} and, for the vector, the acceleration of the Earth and the pulsar \cite{Xue:2021dp}. While including these contributions could lead to stronger constraints, our goal here is to assess the sensitivity of the orbital channel in isolation.

\subsection{Orbital perturbations}
\label{sec:orbital}

We adopt the osculating-elements formalism, treating the dark-matter effect as a perturbation. In this framework, the binary follows an instantaneous Keplerian orbit at each moment, while its orbital elements evolve with time under the perturbation. The standard equations for the evolution of the orbital parameters, given the components of a perturbing force per unit mass in the polar basis $(\hat r,\hat\theta,\hat z)$, are collected in Appendix \ref{app:osculating}. We evaluate them on the unperturbed orbit, to first order in the coupling.

In Appendix \ref{app_orbital}, we collect the expressions for the variations of the five parameters that enter Eqs. \eqref{eq: modelo DD} and \eqref{eq: modelo ELL1}. The scalar expressions follow Refs. \cite{Kus2024, Huxhagen2026}, where the vector case was treated only through the orbital phase; here we complete it with the variations of all five parameters, both in the circular limit and for general eccentricity. For four of the five fitted parameters the perturbation enters the residual algebraically, and the prescription is the one used in those works: the variation $\delta p_\mu(t)$ obtained from the osculating equations is inserted directly into Eq. \eqref{eq:template}. The orbital period is the exception, and requires some care.

The reason is that $P_b$ does not enter the residual directly; instead, it determines the rate at which orbital phase accumulates from a fixed reference epoch:
\begin{equation} 
\Psi'(t) = 2\pi \int_{T_{\rm asc}}^{t} \frac{dt'}{P_b(t')} , \label{eq:phase-def} 
\end{equation}
where $T_{\rm asc}\equiv T_{\rm asc}^{(1)}$ is fixed to its value in the published best-fit solution and is common to all the times of arrival (TOAs) of the given binary. Writing $\bar t = t - T_{\rm asc}$ and $\omega_b = 2\pi/P_b$, the familiar $\Psi' = \omega_b \, \bar t$ is the particular case of a constant period\footnote{The distinction is drawn in Ref. \cite{BlandfordTeukolsky1976} itself, which writes the mean anomaly in terms of a fixed period, its Eq. (2.39), but states that its actual definition is the integral of $a^{-3/2}(t')$ over the elapsed time, its Eq. (2.40).}. In the eccentric basis, $T_{\rm asc}$ is replaced by $T_0\equiv T_0^{(1)}$, $\bar t$ by $\tilde t = t - T_0$, and $\Psi'$ by $\Theta'$.

The timing fit, on the other hand, linearizes the residual about the published solution, which assumes a constant reference period. The corresponding partial derivative is therefore
\begin{equation} \frac{\partial R}{\partial P_b} = \frac{\partial R}{\partial \Psi'} \, \frac{\partial \Psi'}{\partial P_b}=   -\,\frac{\omega_b \, \bar t}{P_b} \, \frac{\partial R}{\partial \Psi'}, \label{eq:pb-column} \end{equation}
and the explicit factor $\bar t$ that it carries corresponds to the phase that a constant offset of the period accumulates over the elapsed time. The perturbation itself follows from Kepler's third law, which ties the period to the semi-major axis and the total mass,
\begin{equation} \frac{\delta P_b(t)}{P_b} = \frac{3}{2}\,\frac{\delta a(t)}{a} - \frac{1}{2}\,\frac{\delta \mathrm{M}_{\rm T}(t)}{\mathrm{M}_{\rm T}} , \label{eq:pb-kepler} \end{equation}
the second term being present only for the scalar couplings, which modulate the masses. Expanding Eq. \eqref{eq:phase-def} to first order in $\delta P_b/P_b$ gives instead
\begin{equation} \delta\Psi'(t) = -\,\frac{\omega_b}{P_b} \int_{T_{\rm asc}}^{t} \delta P_b(t')\,dt' = -\,\frac{\omega_b \, \bar t}{P_b} \, \overline{\delta P_b}(t) , \label{eq:phase-averaged} \end{equation}
where
\begin{equation}
\overline{\delta P_b}(t) \equiv \frac{1}{\bar t}\int_{T_{\rm asc}}^{t} \delta P_b(t')\,dt'.
\end{equation}
For a constant perturbation, $\overline{\delta P_b}(t)=\delta P_b$, and the usual response to an offset in the fitted period is recovered. More generally, the same partial derivative of Eq. \eqref{eq:pb-column} may still be used, provided it is fed the mean of the perturbation over the interval from the reference epoch, rather than its instantaneous value\footnote{The treatment of a secular period change in Ref. \cite{BlandfordTeukolsky1976} is the linear case of Eq. \eqref{eq:phase-averaged}: for $\delta P_b(t') = \dot P_b \, \tilde t'$ the running mean is $\overline{\delta P_b} = \dot P_b \, \tilde t / 2$, and Eq. \eqref{eq:pb-eff} reproduces the factor of one half that accompanies $\dot P_b$ in that model, Eqs. (2.38) and (2.41) of that work. The instantaneous prescription returns the same expression without that factor.},
\begin{equation} \delta P_b(t)\big|_{\rm eff} = \overline{\delta P_b}(t) . \label{eq:pb-eff} \end{equation}

The template is therefore
\begin{equation} h(t) = \sum_{\mu:\, p_\mu \neq P_b} \frac{\partial R}{\partial p_\mu}\, \delta p_\mu(t) \; + \; \frac{\partial R}{\partial P_b}\, \overline{\delta P_b}(t) , \label{eq:template-explicit} \end{equation}
with the partial derivatives read off the design matrix of the published fit\footnote{For the ELL1 basis we restore the term $-\tfrac{3}{2}\,x\,\eta$ of the R\"omer delay, which the model drops as a constant offset \cite{Lange2001}. For a time-dependent $\delta\eta(t)$ it is no longer constant, and without it the cancellation discussed in Sec. \ref{sec:results} for $m\gg\omega_b$ would not occur.} and the $\delta p_\mu(t)$ given in Appendix \ref{app_orbital}.

The perturbing force oscillates at $m$ for the linear scalar and vector couplings and at $2m$ for the quadratic one, while the Keplerian motion is a sum of harmonics of $\omega_b$. When the two frequencies match, $m=n\,\omega_b$ or $2m=n\,\omega_b$ with integer $n$, the resonant terms of Appendix \ref{app_orbital} stop oscillating and the orbital elements drift secularly. Through Eq. \eqref{eq:phase-averaged}, a drift of the period becomes a phase error quadratic in time. Within $|m-n\,\omega_b|\lesssim 1/T_{\rm obs}$ ($2m$ in place of $m$ for the quadratic coupling), the signal exceeds the off-resonance response by several orders of magnitude \cite{Blas2019}.

Which harmonics appear depends on the eccentricity and on the coupling. Eccentric orbits resonate at every harmonic and in every orbital parameter, for the three couplings. On circular orbits, the scalar couplings resonate only at $n=1$ and only in the Laplace parameters $\eta$ and $\kappa$. The vector coupling also makes $x$, $T_{\rm asc}$ and the period resonate at $m=\omega_b$, and the Laplace parameters at $m=2\omega_b$.

\section{Data and Bayesian methodology}
\label{sec:data}

\subsection{The NANOGrav 15 yr data set}
\label{sec:dataset}

We use the NANOGrav 15 yr data set \cite{NANOGrav2023, NANOGrav15data}, the fifth public release of the North American Nanohertz Observatory for Gravitational Waves. It contains observations of $68$ millisecond pulsars (MSPs) taken between 2004 July and 2020 August with the Arecibo Observatory, the Green Bank Telescope and the Very Large Array. Each source is observed roughly monthly, and each epoch combines two widely separated radio bands, so that the dispersive delay of the interstellar medium can be separated from achromatic timing signals.

The release provides two versions of the same observations, differing in how they treat the frequency dependence of the pulse profile. The narrowband version divides the band into frequency channels and measures a separate TOA in each by matching a one-dimensional template, so that a single epoch yields several tens of TOAs; the dispersion measure (DM) is then recovered \textit{a posteriori} as a timing-model parameter, through the DMX model that assigns an independent DM to each observing epoch \cite{Demorest2013}. The wideband version fits the whole receiver band at once against a two-dimensional, frequency-dependent template (a portrait that models the evolution of the pulse shape across the band), and the alignment of the portrait with the data returns a pulse phase and a DM simultaneously \cite{Pennucci2014, Pennucci2019}. Each wideband measurement therefore consists of one TOA per receiver band per epoch, accompanied by a DM value with its own uncertainty, which is passed to the timing model as a datum. This compresses the data set by a factor of $\sim\!33$, from a total of $676{,}465$ narrowband TOAs to $20{,}290$ wideband ones, at little cost in timing precision: wideband and narrowband analyses of the same observations return broadly consistent timing solutions, noise estimates and residual statistics \cite{Alam2021wb}.

We adopt the wideband version throughout, because the ULDM signal is strictly achromatic and enters the residuals only through the Keplerian elements of the orbit. The intra-band frequency structure of the profile carries no information that distinguishes the signal from noise, so collapsing the band into a single TOA plus a DM measurement leaves the template unchanged while removing most of the chromatic parameters from the fit. Our analysis requires one Markov chain per system, per mass and per coupling, and the cost of each likelihood evaluation scales with the number of TOAs, so the smaller wideband data set makes it far less computationally demanding.

Of the $68$ MSPs in the release, $18$ are isolated and $50$ are members of binary systems. Since the couplings we consider act on the residuals only through the orbital motion, our sample consists of the $50$ binaries. We focus on the mass range $10^{-23}$ to $10^{-18}$ eV, motivated by the orbital frequencies of the pulsars in our sample. Their orbital periods span from $1.6$ h to $175.5$ d, corresponding to fundamental resonances $m=\omega_b$ distributed between $2.7\times10^{-22}$ and $7.2\times10^{-19}$ eV.

The sample is also mostly circular: $47$ of the $50$ systems have $e \le 8\times10^{-4}$, as expected for MSPs recycled through a long process of stable mass transfer \cite{Bhattacharya1991}. The three exceptions are PSR J1903$+$0327 ($e = 0.44$), PSR J1946$+$3417 ($e = 0.13$) and PSR J2234$+$0611 ($e = 0.13$). We describe the $47$ near-circular systems in the ELL1 parametrization and the three eccentric ones in the Damour--Deruelle (DD) model. The release is not uniform in this respect: only $26$ of the $47$ near-circular systems are published in ELL1 or ELL1H \cite{FreireWex2010}, the other $21$ being given in eccentric-orbit timing models (DD, BT or DDK \cite{Kopeikin1996}) despite their small eccentricities. We reparametrize those $21$ to ELL1, so that a single circular template applies to all $47$. For them we keep the residuals of the published fit and use the ELL1 model only to build the design matrix and the template, whose variations are taken with respect to its parameters.

None of the $50$ companions is a neutron star: $46$ have minimum masses below $0.6\,M_\odot$, and the four heavier ones are three white dwarfs and one main-sequence star. For the vector coupling we therefore quote limits for $\Delta c = 0.5$, as in previous searches for $B-L$ dark photons \cite{delta_c, Xue:2021dp}, which underestimates $\Delta c$---and is thus conservative---for hydrogen-rich companions.

From the release we take the TOAs, referred to the Solar System barycenter with the DE440 ephemeris \cite{Park2021} as in the published solutions, the fitted timing models, and the single-pulsar noise chains distributed with it, which supply the white-noise parameters of every receiver--backend combination and their wideband DM counterparts. These chains follow the noise model of Ref. \cite{NANOGravNoise}.

\subsection{Noise and signal modeling}
\label{sec:noise}

For a given pulsar we write the time residuals as
\begin{equation}
    \delta\mathbf{t} = \mathbf{h}(\xi,\varrho,\Upsilon,\ldots) + \mathbf{M}\,\boldsymbol\epsilon + \mathbf{F}\,\mathbf{a} + \mathbf{n},
    \label{eq:datamodel}
\end{equation}
where $\mathbf{h}$ is the ULDM delay $h(t)$ of Eq. \eqref{eq:template-explicit}, induced by the time-dependent orbital parameters and evaluated at the arrival times. The other three terms collect everything we do not wish to measure. $\mathbf{M}$ is the timing-model design matrix, whose columns are the partial derivatives of the residuals with respect to each fitted timing parameter, evaluated at the published solution; $\boldsymbol\epsilon$ holds small offsets to those parameters. These are the spin frequency and its derivatives, the astrometric parameters, the five orbital parameters, and one dispersion-measure coefficient per observing epoch, the last of these being by far the most numerous. $\mathbf{F}$, which describes the achromatic red noise, is a Fourier design matrix of sines and cosines at the frequencies $f_k = k/T_{\rm obs}$, with $k = 1,\ldots,30$ and $T_{\rm obs}$ the observing span of the pulsar, and $\mathbf{a}$ holds their coefficients. The last term, $\mathbf{n}$, is the white noise: a zero-mean Gaussian vector with one entry per arrival time, whose covariance we write as $\mathbf{N} = \langle \mathbf{n}\,\mathbf{n}^{T}\rangle$. Throughout we follow the wideband formulation of Ref. \cite{Alam2021wb}, whose Appendix B derives the likelihood we build in this section.

The white-noise covariance is diagonal, and each entry rescales the uncertainty $\sigma_i$ of TOA $i$ through the EFAC $\mathcal{E}_j$ and the EQUAD $\mathcal{Q}_j$ of the receiver--backend combination $j$ that produced it. We use the convention of the NANOGrav 15 yr noise model \cite{NANOGravNoise},
\begin{equation}
    N_{ii} = \mathcal{E}_{j}^{2}\,\big(\sigma_{i}^{2} + \mathcal{Q}_{j}^{2}\big).
    \label{eq:whitenoise}
\end{equation}
The EFAC absorbs a misestimation of the template-fitting error, and the EQUAD an additional variance uncorrelated between epochs. No ECORR term appears, since a wideband TOA is already the single measurement of its band and epoch.

We model the achromatic red noise as a stationary Gaussian process with a power-law spectral density. Red spin noise is expected in millisecond pulsars \cite{ShannonCordes2010} and is detected in several pulsars of this data set. This parametrization is the one adopted throughout PTA analyses \cite{Lentati2013},
\begin{equation}
    S_{\rm red}(f) = \frac{A_{\rm red}^{2}}{12\pi^{2}}\left(\frac{f}{f_{\rm yr}}\right)^{-\gamma_{\rm red}} f_{\rm yr}^{-3},
    \label{eq:redpsd}
\end{equation}
with $f_{\rm yr} = 1\,{\rm yr}^{-1}$ the reference frequency at which $A_{\rm red}$ is defined, following the characteristic-strain convention that fixes the factor $12\pi^{2}$ \cite{vanHaasterenVallisneri2014}. As in the noise analysis released with the data set, each of the $30$ harmonics contributes a sine and a cosine column, so $\mathbf{a}$ has $60$ entries. Their prior is Gaussian, zero-mean and diagonal in this basis: the two coefficients of harmonic $k$ share the variance $S_{\rm red}(f_k)\,\Delta f$, with $\Delta f = 1/T_{\rm obs}$ the spacing of the Fourier basis.

The same data set shows evidence for a stochastic gravitational-wave background. It is a red process common to all pulsars, with amplitude $A\simeq2.4\times10^{-15}$ at $f_{\rm yr}$ for $\gamma=13/3$ \cite{nano_GW}, and an array identifies it through the Hellings--Downs correlation between pulsars \cite{HellingsDowns1983}. Within one pulsar, the background adds a second power law to the intrinsic spin noise, which the single-pulsar noise chains released with the data set already absorb into the individual red noise. Our priors on $A_{\rm red}$ and $\gamma_{\rm red}$ contain its values, so the single power law of Eq. \eqref{eq:redpsd} can approximately describe both.

The entries of $\boldsymbol\epsilon$ are treated in two ways. Almost all of them carry a prior so broad that the likelihood is insensitive to it, so the data alone determine them. The exceptions are the DMX coefficients, the epoch-by-epoch offsets of the dispersion measure. In the wideband data every TOA comes with its own DM measurement, and we use the measurements of each epoch as a Gaussian prior on the DMX offset of that epoch, with mean and width set by the measured values and their uncertainties. Before doing so we subtract from the measurements of each receiver a constant offset, fitted in the published solution as the timing-model parameter DMJUMP, which removes the difference in DM reference between the templates of different receivers. The mean of the prior is then known, so we remove from the residuals the delay it implies and are left with a zero-mean prior. The global DM is held fixed, so the measurements of an epoch constrain only that epoch's DMX coefficient. As a result, each DMX offset can move only within the uncertainty of its own DM measurement, and cannot absorb an achromatic signal such as ours.

We hold the white-noise parameters $(\mathcal{E}_j,\mathcal{Q}_j)$ and their dispersion-measure counterparts $(\mathcal{E}_{{\rm DM},j},\mathcal{Q}_{{\rm DM},j})$ fixed at the maximum \textit{a posteriori} values of the wideband single-pulsar noise chains released with the data set. They are constrained by the high-frequency scatter of the residuals, which no ULDM signal in our mass range can mimic, so sampling them again would multiply the cost of every chain for no gain. The red-noise hyperparameters, by contrast, are sampled. At the lower end of our mass range the signal is slow compared with the observing span and partly degenerate with red noise, and fixing $A_{\rm red}$ and $\gamma_{\rm red}$ at values fitted in the absence of a signal would make the limit optimistic precisely there.

One timing parameter is treated differently. The published solutions of nine of our binaries fit a secular derivative of the period, $\dot P_b$, which enters the phase quadratically in time. On resonance the field produces the same quadratic phase over the observing span, so a free $\dot P_b$ absorbs part of the resonant signal close to the center of each resonance. We hold $\dot P_b$ at its published value in every chain, for all couplings and masses. This amounts to assuming that the measured derivative is accounted for by gravitational-wave emission \cite{Peters1964}, the Shklovskii effect \cite{Shklovskii1970} and the Galactic acceleration \cite{DamourTaylor1991, Chakrabarti2021}, the same assumption under which seven of these systems are used to map the Galactic acceleration in Ref. \cite{Moran2024}. We show in Appendix \ref{app:pbdot} that marginalizing $\dot P_b$ instead changes the limit only within a narrow band at the center of each resonance. The other secular derivatives remain free and are marginalized.

The signal brings its own parameters. Besides the coupling $\xi \in \{\alpha,\beta,g\}$, the template depends on the local realization of the field: the amplitude $\varrho$ and phase $\Upsilon$, sampled from their Rayleigh and uniform priors, and in the vector case the direction $\hat n$, sampled isotropically. The vector template also depends on the semi-major axis $a$ of the relative orbit and on the inclination $\iota$, which we fix for each system from the published masses, inclination and mass function as described in Appendix \ref{app:sample}. The masses enter through $a\propto \mathrm{M}_{\rm T}^{1/3}$, so their uncertainty has little effect on the template.

The ULDM mass is not sampled. We fix it at each point of a grid and repeat the analysis. The prior on the coupling is uniform in $\xi$ for the upper limits, and uniform in $\log_{10}\xi$ over the same range for the injection tests of Appendix \ref{app:injection}. Table \ref{tab:priors} lists all the parameters of the single-pulsar analysis and their priors.

\begin{table}
\centering
\footnotesize
\caption{Parameters of the single-pulsar analysis. $\mathcal{U}$ denotes a uniform and $\mathcal{N}$ a Gaussian prior; the parameters with a Gaussian prior are marginalized analytically. For the couplings we give the range over which the prior is uniform in $\xi$ for the upper limits and uniform in $\log_{10}\xi$ for the injection tests.}
\label{tab:priors}
\begin{tabular}{ll}
\toprule
Parameter & Prior \\
\midrule
$\alpha$ [GeV$^{-1}$] & $\mathcal{U}[10^{-26},10^{-10}]$ \\
$\beta$ [GeV$^{-2}$] & $\mathcal{U}[10^{-44},10^{-12}]$ \\
$g$ & $\mathcal{U}[10^{-30},10^{-14}]$ \\
$\varrho$ & $2\varrho\,e^{-\varrho^{2}}$ \\
$\Upsilon$ & $\mathcal{U}[0,2\pi)$ \\
$\cos\vartheta$ (vector) & $\mathcal{U}[-1,1]$ \\
$\varphi$ (vector) & $\mathcal{U}[0,2\pi)$ \\
$m$ & fixed on a grid \\
\midrule
$\log_{10}A_{\rm red}$ & $\mathcal{U}[-20,-11]$ \\
$\gamma_{\rm red}$ & $\mathcal{U}[0,10]$ \\
Fourier coefficients $a_k$ & $\mathcal{N}\big(0,\,S_{\rm red}(f_k)/T_{\rm obs}\big)$ \\
\midrule
EFAC, EQUAD & fixed at MAP \\
DMEFAC, DMEQUAD & fixed at MAP \\
\midrule
DMX offsets & $\mathcal{N}\big(0,\,\sigma_{\rm DMX}^{2}\big)$ \\
$\dot P_b$ & fixed at published value \\
other timing offsets & flat \\
\bottomrule
\end{tabular}
\end{table}

\subsection{Bayesian methodology}
\label{sec:bayes}

TOAs and timing models are handled with \texttt{PINT} \cite{Luo_2021}, and the likelihood is built with \texttt{ENTERPRISE} \cite{enterprise} from the white noise of Eq. \eqref{eq:whitenoise}, the red noise of Eq. \eqref{eq:redpsd} and the wideband timing model. Conditional on the linear parameters $\boldsymbol\epsilon$ and $\mathbf{a}$, $\delta\mathbf{t} - \mathbf{h} - \mathbf{M}\,\boldsymbol\epsilon - \mathbf{F}\,\mathbf{a}$ has a Gaussian distribution of zero mean, with covariance $\mathbf{N}$.

We do not sample those parameters; we marginalize over them analytically \cite{vanHaasteren2013}. Collecting them into $\mathbf{b} = (\boldsymbol\epsilon,\mathbf{a})$ and writing $\mathbf{T} = [\,\mathbf{M}\;\mathbf{F}\,]$, a zero-mean Gaussian prior of covariance $\mathbf{B}$ turns the integral over $\mathbf{b}$ into a Gaussian one, and what is left is a likelihood of the form
\begin{equation}
    \mathcal{L} = \frac{\exp\!\big[-\tfrac{1}{2}\,(\delta\mathbf{t} - \mathbf{h})^{T}\,\mathbf{C}^{-1}\,(\delta\mathbf{t} - \mathbf{h})\big]}{\sqrt{\det\!\big(2\pi\mathbf{C}\big)}},
    \label{eq:likelihood}
\end{equation}
in which the marginalized terms now appear in the covariance,
\begin{equation}
    \mathbf{C} = \mathbf{N} + \mathbf{T}\,\mathbf{B}\,\mathbf{T}^{T},
    \qquad
    \mathbf{B} = \mathrm{diag}\big(\mathbf{B}_{\rm tm},\,\boldsymbol\Sigma_{\rm red}\big),
    \label{eq:covarianza}
\end{equation}
with $\mathbf{B}_{\rm tm}$ the covariance of the timing-model offsets and $\boldsymbol\Sigma_{\rm red}$ that of the red-noise coefficients. 

Both blocks are diagonal and hold the priors described above: $[\boldsymbol\Sigma_{\rm red}]_{kk} = S_{\rm red}(f_k)\,\Delta f$, with each harmonic entering twice, $\sigma_{\rm DMX}^{2}$ on each DMX coefficient, and a variance large enough to act as a flat prior on every other timing parameter. Since $\boldsymbol\Sigma_{\rm red}$ depends on the sampled $(A_{\rm red},\gamma_{\rm red})$ while $\mathbf{N}$ and the DMX entries of $\mathbf{B}_{\rm tm}$ are held fixed, $\mathbf{C}$ is rebuilt at every step of the chain.

What we want is the posterior on the coupling alone, with every other parameter integrated out. Let $\lambda_i = (\varrho_i,\Upsilon_i,\hat n_i,\log_{10}A_{{\rm red},i},\gamma_{{\rm red},i})$ denote the nuisance parameters of pulsar $i$, where $\hat n_i$ appears only in the vector case, and let $\Lambda=(\lambda_1,\ldots,\lambda_{N_p})$ and $\delta\mathbf{t}=(\delta\mathbf{t}_1,\ldots,\delta\mathbf{t}_{N_p})$ collect the parameters and residuals of the whole array. The target is then
\begin{equation}
    p(\xi \,|\, \delta\mathbf{t}) \propto \pi(\xi) \int \mathcal{L}(\delta\mathbf{t} \,|\, \xi, \Lambda)\, \pi(\Lambda)\, \mathrm{d}\Lambda .
\end{equation}

In the uncorrelated regime, each binary sees an independent realization of the field, so that no two systems share an amplitude, a phase or a direction. The residuals of different pulsars are then independent, the likelihood factorizes as $\mathcal{L} = \prod_i \mathcal{L}_i$, and so does the prior, $\pi(\Lambda)=\prod_i\pi(\lambda_i)$. Since each $\lambda_i$ appears in one factor only, the integral splits accordingly,
\begin{equation}
    p(\xi \,|\, \delta\mathbf{t}) \propto \pi(\xi) \prod_{i=1}^{N_p} Z_i(\xi),
    \label{eq:factorization}
\end{equation}
where
\begin{equation}
    Z_i(\xi) = \int \mathcal{L}_i(\delta\mathbf{t}_i \,|\, \xi,\lambda_i)\, \pi(\lambda_i)\, \mathrm{d}\lambda_i
\end{equation}
is the marginal likelihood of pulsar $i$ as a function of the coupling.

Besides the independence of the field realizations, the factorization neglects the effects of a stochastic gravitational-wave background and of the ultralight dark matter field on the propagation of the pulses from the pulsar to Earth, which have already been analyzed with PTAs \cite{nano_GW, EPTA2023, Reardon2023, Xu2023, KhmelnitskyRubakov2014, Porayko:2018sfa, Smarra:2023rjx, Afzal2023}. Their correlation among pulsars is what distinguishes these effects from the intrinsic red noise of each pulsar. Here we focus on the orbital effect, which according to previous forecasts \cite{Kus2024, Huxhagen2026} dominates at higher masses of the field, and in Sec. \ref{sec:results} we compare our limits in the orbital channel with those obtained from PTAs at lower masses. In the mass range we consider, the orbital signal is not coherent across pulsars: it is produced by the local field acting on the orbit of each binary, and most of its power lies near the orbital harmonics, $f\gtrsim1/P_b\gtrsim6\times10^{-8}\,$Hz, rather than at the lowest frequencies. We therefore expect neglecting these correlations to have a small effect on the limits, while a joint analysis of the whole array with a correlated common process would multiply the computational cost of the analysis.

Equation \eqref{eq:factorization} is what licenses one chain per system, rather than a single array-wide chain. We run $N_p$ independent samplers. Each explores $(\log_{10}\xi,\lambda_i)$ in five dimensions for the scalar couplings, or seven for the vector, and returns the single-pulsar posterior $p_i(\xi) \propto \pi(\xi)\,Z_i(\xi)$. Dividing out the prior recovers the marginal likelihood up to a constant, $Z_i(\xi) \propto p_i(\xi)/\pi(\xi)$, which we estimate from a histogram of the $\log_{10}\xi$ samples. 
Chains are run with \texttt{PTMCMCSampler} \cite{PTMCMC}, and the scheme is trivially parallel.

The chains are then combined at fixed mass to set an upper limit on the coupling, with $\pi(\xi)$ uniform in $\xi$. Taking logarithms in Eq. \eqref{eq:factorization} gives
\begin{equation}
    \ln p(\xi \,|\, \delta\mathbf{t}) = \ln \pi(\xi) + \sum_{i=1}^{N_p} \ln Z_i(\xi) + \mathrm{constant},
    \label{eq:combinacion}
\end{equation}
and we define the upper bound $\xi_{95}$ as the value that leaves $95\%$ of this posterior below it,
\begin{equation}
    \int_{0}^{\xi_{95}} p(\xi \,|\, \delta\mathbf{t})\, \mathrm{d}\xi = 0.95 .
    \label{eq:bound}
\end{equation}
The $\ln Z_i$ of all systems are summed on a common grid in $\xi$. The prior ranges of Table \ref{tab:priors} are wide enough not to affect the results: their upper edges lie at least two orders of magnitude above the combined bound, and their lower edges well below the couplings any system can detect.

\section{Results}
\label{sec:results}
\begin{figure*}
    \centering
    \includegraphics[width=0.49\textwidth]{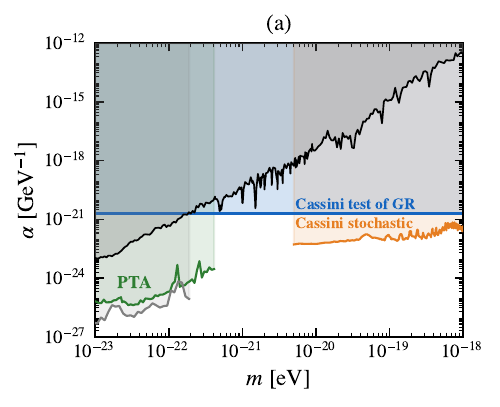}\hfill
    \includegraphics[width=0.49\textwidth]{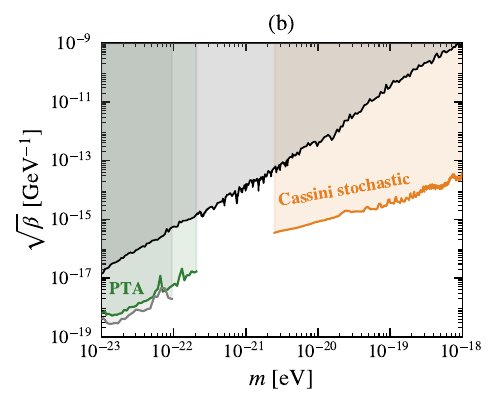}
    \caption{Constraints at $95\%$ credibility as a function of mass on (a) the coupling $\alpha$ of linearly coupled and (b) $\sqrt{\beta}$ of quadratically coupled scalar dark matter. Shaded regions indicate the parameter space excluded by the Cassini test of general relativity \cite{Bertotti:2003rm} (blue), by two pulsar timing array searches for the purely gravitational ULDM signal \cite{Porayko:2018sfa, Smarra:2023rjx} (green and gray, respectively), whose amplitude limits we recast onto the couplings as in Refs. \cite{Blas2017Resonates, Huxhagen2026}, and by the Cassini search for a stochastic gravitational-wave background \cite{Armstrong:2003hb} (orange). Solar System constraints from tests of general relativity have not yet been derived for the quadratic coupling.}
    \label{fig:constraints-scalar}
\end{figure*}

We present $95\%$ credible upper limits on the three couplings considered in this work. Figure \hyperref[fig:constraints-scalar]{\ref*{fig:constraints-scalar}(a)} shows the limit for linearly coupled scalar ULDM. Away from resonances it weakens with mass from $10^{-23}\,\mathrm{GeV}^{-1}$ at the lower end of our range to a few times $10^{-13}\,\mathrm{GeV}^{-1}$ at the upper end. Below $m\simeq2\times10^{-22}\,$eV it lies below the Cassini bound, but there the recast PTA limits are considerably stronger. On top of this, the orbital resonances produce two kinds of features. The deepest ones, around $10^{-21}\,$eV, come from the eccentric systems; they fall within the expected orders of magnitude and, led by the low harmonics of PSR J1903$+$0327, bring the limit close to the Cassini bound, although other bounds remain stronger over the whole mass range. At higher masses, shallower dips appear at the orbital frequencies of circular binaries. These are the $n=1$ resonances of the Laplace parameters $\eta$ and $\kappa$. Up to $10^{-20}\,$eV the limit agrees with the estimates of Ref. \cite{Kus2024}, and at higher masses it is weaker. That estimate follows the projected semi-major axis on its own. When the field oscillates faster than the orbit, all the orbital elements oscillate, and their contributions to the arrival times largely cancel, leaving a much smaller displacement of the pulsar than the variation of $x$ alone would suggest.

Figure \hyperref[fig:constraints-scalar]{\ref*{fig:constraints-scalar}(b)} presents our results for the quadratic coupling. While limits from pulsar timing arrays and the Cassini search for stochastic gravitational waves are stronger than our bound in their respective mass ranges, we report the best current constraints on a universal quadratic coupling in the range from $2\times10^{-22}\,\mathrm{eV}$ to $2\times10^{-21}\,\mathrm{eV}$. The resonant peaks characteristic of eccentric systems are visible, although, unlike in the linear case, they improve the off-resonance combined bound by less than an order of magnitude. Up to $10^{-20}\,$eV the bound agrees with the sensitivity estimates of Ref. \cite{Huxhagen2026}, and at higher masses it is weaker, for the same reason as in the linear case. These limits treat the neutron star as a point particle in the external field. Inside matter of density $\rho$, a quadratically coupled field acquires an effective mass $m_{\rm eff}^2\simeq m^2+\beta\rho$, which can screen it inside the star and weaken the limits by a few orders of magnitude \cite{Blas2017Resonates, Blas2019}.

\begin{figure}
    \centering
    \includegraphics[width=0.99\linewidth]{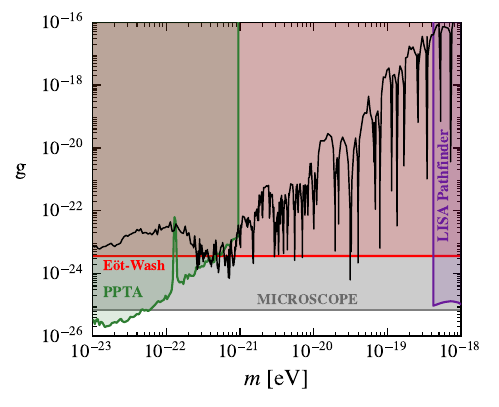}
    \caption{Constraints on the coupling $g$ to the $B-L$ current at $95\%$ credibility as a function of mass for vector dark matter. Shaded regions indicate the parameter space excluded by the E\"ot-Wash torsion-balance test of the equivalence principle \cite{Wagner2012, AxionLimits} (red), MICROSCOPE \cite{Amaral:2024vwd} (gray), LISA Pathfinder \cite{Frerick:2023dm} (purple), and the Parkes Pulsar Timing Array \cite{Xue:2021dp} (green).}
    \label{fig:constraint-g}
\end{figure}

Lastly, we present our constraints on vector ULDM coupled to baryon number minus lepton number in Fig. \ref{fig:constraint-g}. MICROSCOPE \cite{Amaral:2024vwd} gives the strongest bound at every mass in our range, about an order of magnitude or more below ours. Between $3\times10^{-22}$ and $10^{-21}\,$eV our limit is comparable to those of the Parkes Pulsar Timing Array \cite{Xue:2021dp} and the E\"ot-Wash test of the equivalence principle \cite{Wagner2012}, and at the orbital resonances in that interval it falls below both. At higher masses, beyond the reach of the PPTA search, it approaches or beats the E\"ot-Wash bound only at isolated resonances. Above $4\times10^{-19}\,$eV, LISA Pathfinder \cite{Frerick:2023dm} is stronger by more than four orders of magnitude. Compared with the forecast of Ref. \cite{Huxhagen2026}, evaluated for the same charge difference\footnote{That forecast was quoted for $\Delta c=0.1$; here we use $\Delta c=0.5$, and the comparison is made at that value.}, the off-resonance limit is weaker above $10^{-20}\,$eV, for the same reason as in the linear case. Below $10^{-21}\,$eV it stays within half an order of magnitude of the forecast, and at the lowest masses it is stronger. This is due to considering the variations of all orbital parameters, while the forecast follows the orbital phase alone, which makes our analysis more sensitive.

On resonance the peaks are one to two orders of magnitude shallower than forecast, as a consequence of marginalizing over the direction of the field. There, the signal is dominated by a secular drift of the period, and on circular orbits its amplitude is proportional to $\varrho\sin\vartheta\cos(\Upsilon_{\rm asc}+\varphi_{\rm asc})$, Eq. \eqref{eq:vector_resonance}, and an analogous combination of $\varphi$ and $\Upsilon_0$ appears on eccentric orbits. A field nearly aligned with the orbital angular momentum leaves the period unchanged and only tilts the orbit, to which timing is far less sensitive, and a field with an in-plane component can still produce little drift if its phase and azimuth combine unfavorably. In the scalar case only the phase can suppress the resonant drift; in the vector case the direction adds a second factor, which makes such configurations more probable \textit{a priori}. They do not affect the typical sensitivity of the forecast, computed from the median over realizations, but they dominate the tail of the marginalized posterior and therefore the credible limit.

\section{Conclusions}
\label{sec:conclusions}

We have used the arrival times of the $50$ binaries of the NANOGrav 15 yr data set to constrain ultralight dark matter coupled directly to ordinary matter through its effect on their orbits, between $10^{-23}$ and $10^{-18}\,$eV. We set $95\%$ credible limits on a linear and a quadratic universal coupling of a scalar field, and on the coupling of a vector field to $B-L$. These limits cover a continuous range of masses in the orbital channel, which had so far been studied through forecasts.

For the linear coupling, the binaries improve on the Cassini bound only at low masses, where the recast PTA limits are stronger, and elsewhere approach it only at the resonances of the eccentric systems. For a universal quadratic coupling they fill the window between the masses reached by pulsar timing array searches and those of the Cassini search for gravitational waves, where, between $2\times10^{-22}$ and $2\times10^{-21}\,$eV, they give the strongest limit available. For the $B-L$ vector, MICROSCOPE remains at least an order of magnitude stronger at every mass, and the binaries are competitive with the PPTA and E\"ot-Wash bounds below $10^{-21}\,$eV and at isolated resonances above.

Away from resonance the limits follow the forecasts of Refs. \cite{Kus2024, Huxhagen2026} at low masses and are weaker above $10^{-20}\,$eV, where the contributions of the orbital elements to the arrival times largely cancel. On resonance the scalar features fall within the expected orders of magnitude, while the vector ones are one to two orders of magnitude shallower than forecast. A credible limit is set by the realizations of the field least favorable to detection, whereas a forecast describes a typical one, and the difference between the two is largest when the orientation of the field matters, as it does for a vector.

A consistent treatment of the screening of a quadratically coupled field inside the pulsar requires solving for the field together with the structure of the star. Building on studies of the field profile in and around dense bodies \cite{Hees2018, gan2026probing, burrage2026impact}, this can be done numerically, as it has been for the response of neutron stars to a scalar field in scalar--tensor theories \cite{KuntzBarausse}; we leave it for future work.

More generally, the analysis presented here applies to any field that perturbs the orbit, since only the variations $\delta p_\mu(t)$ change from one coupling to another, and it can be extended to other models such as spin-2 ultralight dark matter.

The orbital channel uses the same data as the searches for an oscillation in the arrival times themselves, and extends them to masses where the field oscillates faster than the observing cadence. On resonance the orbital elements drift secularly, so a longer observing span improves the resonant limits faster than the off-resonance one. The largest gains would come from eccentric binaries with long and precise timing, since they resonate at every harmonic and in every orbital parameter, and only three of our systems are eccentric. Longer and combined data sets from other arrays, and in the longer term the Square Kilometre Array \cite{SKA_Science_2015}, will add both span and systems.

\section*{Data availability statement}
This work uses the wideband data of the NANOGrav 15 yr data set, which is publicly available \cite{NANOGrav15data}. The single-pulsar posterior histograms, the analysis pipelines and the final exclusion curves are available in a Zenodo repository \cite{HuxhagenZenodo2026}.

\section*{Acknowledgments}

This work has been supported by CONICET and UBA. F.H. acknowledges partial financial support from CNES. The computational resources used in this work were provided (in part) by the HPC center DIRAC, funded by Instituto de Física de Buenos Aires (UBA-CONICET) and part of the SNCAD-MinCyT initiative, Argentina. We thank the NANOGrav collaboration for making the 15 yr data set publicly available. During the development of the code, we used Claude Opus 5.5 to check the implemented formulas against our analytic calculation in Mathematica, to help identify and correct implementation errors, and to prepare its public version. All suggested changes were reviewed by the authors.

\appendix
\section{Osculating equations}
\label{app:osculating}

A perturbing force acting on a binary is most naturally described through the instantaneous Keplerian orbit that would be followed if the perturbation were switched off. Six osculating elements specify that orbit \cite{turner1979}: the semi-major axis $a$ of the relative orbit, the eccentricity $e$, the inclination $\iota$, the longitude of the ascending node $\Omega$, the argument of periastron $\omega$, measured from the node, and the phase element $\epsilon_1$ defined below. Their rates of change follow from the components $(F_r,F_\theta,F_z)$ of the perturbing force per unit mass along the radial, tangential and orbit-normal directions of Fig.~\ref{fig: orbita kepleriana}. Writing $\sigma\equiv\sqrt{1-e^2}$ to keep the expressions compact,
\begin{subequations}
\label{eq:gauss_equations}
\begin{flalign}
    & \dot{a} = \frac{2}{\omega_b} \left\{ \frac{F_r \, e \sin\theta}{\sigma} + \frac{a \, \sigma}{r} \, F_\theta \right\} && \label{eq:gauss_a} \\
    & \dot{e} = \frac{\sigma}{a \, \omega_b} \left\{ F_r \sin\theta + F_\theta \left( \cos\theta + \cos E \right) \right\} && \label{eq:gauss_e} \\
    & \dot{\iota} = \frac{r \cos(\theta + \omega)}{a^2 \omega_b \, \sigma} \, F_z && \label{eq:gauss_iota} \\
    & \dot{\Omega} = \frac{r \sin(\theta + \omega)}{a^2 \omega_b \, \sigma \sin \iota} \, F_z && \label{eq:gauss_Omega} \\
    & \dot{\omega} = \frac{\sigma}{a \, e \, \omega_b} \left\{ - F_r \cos\theta + F_\theta \sin\theta \left[1 + \frac{r}{a \sigma^2} \right] \right\} && \nonumber \\
    & \hspace{25pt} - \cos\iota \, \dot{\Omega} && \label{eq:gauss_omega} \\
    & \dot{\epsilon}_1 = -\frac{2 r}{a^2 \omega_b} F_r + \left( 1 - \sigma \right) \dot{\varpi}
    + \sigma \left( 1 - \cos\iota \right) \dot{\Omega} && \label{eq:gauss_epsilon}
\end{flalign}
\end{subequations}
\noindent Here $r$ is the separation, $\theta$ the true anomaly, that is, the polar angle of Fig.~\ref{fig: orbita kepleriana} measured from periastron, $\varpi = \omega + \Omega$ the longitude of periastron, and $E$ the eccentric anomaly, fixed by
\begin{equation}
e \, \cos E = 1 - \frac{r}{a}.
\end{equation}
The phase element is
\begin{equation}
    \epsilon_1 = \Theta' + \varpi - \int_{T_0}^t \omega_b(t') \, dt',
    \label{eq:eps1_def}
\end{equation}
with $\Theta'$ the orbital phase measured from periastron and $\omega_b(t')$ the orbital angular frequency of the osculating orbit at time $t'$, which varies as the field perturbs the period. For an unperturbed orbit $\Theta'=\omega_b(t-T_0)$, and $\epsilon_1$ reduces to the constant $\varpi$.

The last term of Eq.~\eqref{eq:gauss_omega} appears because $\omega$ is measured from the ascending node, which moves when the force has a component out of the orbital plane. For the scalar couplings $F_z = 0$, so $\iota$ and $\Omega$ stay constant and $\dot\varpi=\dot\omega$; for the vector coupling all six elements evolve. We evaluate the right-hand sides of Eqs.~\eqref{eq:gauss_equations} on the unperturbed orbit, which is consistent to first order in the coupling and is the approximation under which every expression in this work is derived.

Eccentric motion has no closed form in time: the anomalies are related to the mean anomaly $\ell = \omega_b \, \tilde{t}$, with $\tilde{t} = t - T_0$ measured from periastron passage, through Kepler's equation $\ell = E - e\sin E$. The standard route is therefore to expand the unperturbed motion in a Fourier series in $\ell$, whose coefficients are Bessel functions of the first kind of order $n$ evaluated at $ne$ \cite{bessel}. Abbreviating $J_n \equiv J_n(ne)$ and $J_n' \equiv J_n'(ne)$, with the prime denoting the derivative with respect to the argument, the expansions we need are
\begin{subequations}
\label{eq:bessel_series}
\begin{flalign}
    & \frac{a}{r} = 1 + 2 \sum_{n=1}^{\infty} J_n \cos(n \ell) && \label{eq:series_ar} \\
    & \frac{r}{a} = 1 + \frac{e^2}{2} - 2e \sum_{n=1}^{\infty} \frac{J_n'}{n} \cos(n \ell) && \label{eq:series_ra} \\[10 pt]
     & \cos\theta = -e + \frac{2 \sigma^2}{e} \sum_{n=1}^{\infty} J_n \cos(n \ell) && \label{eq:series_costheta} \\
    & \sin\theta = 2 \sigma \sum_{n=1}^{\infty} J_n' \sin(n \ell) && \label{eq:series_sintheta} \\
    & \frac{r}{a} \cos\theta = -\frac{3e}{2} + 2 \sum_{n=1}^{\infty} \frac{J_n'}{n} \cos(n \ell) && \label{eq:series_racos} \\
    & \frac{r}{a} \sin\theta = \frac{2 \sigma}{e} \sum_{n=1}^{\infty} \frac{J_n}{n} \sin(n \ell) && \label{eq:series_rasin} \\
    & \frac{a^2}{r^2} \cos\theta = \sum_{n=1}^{\infty} 2n \, J_n' \cos(n \ell) && \label{eq:series_ar2cos} \\
    & \frac{a^2}{r^2} \sin\theta = \frac{\sigma}{e} \sum_{n=1}^{\infty} 2n \, J_n \sin(n \ell) && \label{eq:series_ar2sin}
\end{flalign}
\end{subequations}
\onecolumngrid
\raggedbottom
\section{Perturbed orbital parameters}
\label{app_orbital}

The osculating variations of $x$, $e$ and $\omega$ (or $\eta$ and $\kappa$ for circular orbits) enter the timing residual directly. The orbital phase does not: the timing model instead fits an orbital period and a reference epoch. In the eccentric basis the phase $\Theta'$ is measured from periastron, and its unperturbed value is the mean anomaly $\ell=\omega_b\tilde t$; in the circular basis the phase $\Psi'$ is measured from the ascending node, and its unperturbed value is $\omega_b\bar t$. The partial derivatives with respect to the fitted parameters are
\begin{equation} \frac{\partial \Theta'}{\partial T_0} = -\omega_b , \qquad \frac{\partial \Theta'}{\partial P_b} = -\frac{\omega_b\tilde t}{P_b} , \label{eq:phase-partials} \end{equation}
and likewise for $\Psi'$, with $T_{\rm asc}$ and $\bar t$ in place of $T_0$ and $\tilde t$. To split the phase perturbation between the period and the epoch, we write the perturbed mean motion as $\omega_b[1-\delta P_b(t)/P_b]$. Since $\dot\Theta'=\omega_b[1-\delta P_b(t)/P_b]+\dot\epsilon_1-\dot\varpi$, with $\epsilon_1$ defined in Eq. \eqref{eq:eps1_def}, and $\Psi'=\Theta'+\omega$,
\begin{equation} \delta\dot\Theta' = \big(\dot\epsilon_1 - \dot\varpi\big) - \omega_b\,\frac{\delta P_b(t)}{P_b} , \qquad \delta\dot\Psi' = \big(\dot\epsilon_1 - \dot\Omega\big) - \omega_b\,\frac{\delta P_b(t)}{P_b} . \label{eq:phase-split} \end{equation}
The period absorbs the second term through the running mean of Eq. \eqref{eq:pb-eff}. The first term is a phase offset absorbed by the epoch of periastron or of the ascending node,
\begin{subequations}\label{eq:epoch-channel}
\begin{align}
\delta T_0
&=-\frac{1}{\omega_b}\int_{0}^{\tilde t}
  (\dot\epsilon_1-\dot\varpi)\,d\tilde t',\\
\delta T_{\rm asc}
&=-\frac{1}{\omega_b}\int_{0}^{\bar t}
  (\dot\epsilon_1-\dot\Omega)\,d\bar t'.
\end{align}
\end{subequations}
Together, the epoch and the running mean of the period reproduce the full phase perturbation. Below we list only the five variations that enter the template of Eq. \eqref{eq:template-explicit} for each coupling and basis.

Throughout, $a$ is the semi-major axis of the relative orbit and $a_1$ that of the pulsar, $\Phi_0=\varrho\sqrt{2\rho_{\rm DM}}/m$ is the amplitude of the scalar field, $F_0$ that of the vector force of Eq. \eqref{eq:fuerza vectorial polar}, and $J_n\equiv J_n(ne)$ and $J_n'\equiv J_n'(ne)$ as in Appendix \ref{app:osculating}. The circular basis uses
\begin{equation*}
\bar t=t-T_{\rm asc},\qquad \Upsilon_{\rm asc}=mT_{\rm asc}+\Upsilon,\qquad \eta=e\sin\omega,\qquad \kappa=e\cos\omega ,
\end{equation*}
and the eccentric basis
\begin{equation*}
\tilde t=t-T_0,\qquad \ell=\omega_b\tilde t,\qquad \Upsilon_0=mT_0+\Upsilon,\qquad \sigma=\sqrt{1-e^2} .
\end{equation*}
In both, $mt+\Upsilon=m\bar t+\Upsilon_{\rm asc}=m\tilde t+\Upsilon_0$. Since the rates are evaluated on the unperturbed orbit, the circular expressions contain the orbital phase only through its zeroth-order value $\omega_b\bar t$, as the eccentric ones contain it through $\ell$. For the period we give $\overline{\delta P_b}/P_b$, the running mean of Eq. \eqref{eq:pb-eff} taken from the reference epoch of each basis.

For eccentric orbits, each rate is the oscillation of the field times harmonics of $\ell$, and integrating it from periastron passage, $\tilde t=0$, gives the primitives $C^\pm_{n,q}$ and $S^\pm_{n,q}$, with $q=1$ for the linear and vector couplings and $q=2$ for the quadratic one. Together with the coefficients $A_n$ and $B_n$, which appear in both scalar expressions for $\delta T_0$, they are
\begin{subequations}\label{eq:ecc_primitives}
\begin{align}
C_{n,q}^\pm
&=\frac{\cos(q\Upsilon_0)
-\cos[(n\omega_b\pm qm)\tilde t\pm q\Upsilon_0]}
{n\omega_b\pm qm},\\
S_{n,q}^\pm
&=\frac{\sin[(n\omega_b\pm qm)\tilde t\pm q\Upsilon_0]
\mp\sin(q\Upsilon_0)}{n\omega_b\pm qm},\\
A_n
&=4\omega_bJ_n-\frac{2n\omega_b\sigma^2}{e}J_n',\\
B_n
&=\frac{4}{n}J_n+\frac{4\sigma^2}{e}J_n'.
\end{align}
\end{subequations}

\subsection{Linearly coupled scalar ULDM}
\subsubsection{Circular orbits}

The five variations entering the circular timing template are
\begin{subequations}
\label{eq:scalar_linear_circular}
\begin{align}
\frac{\delta x}{x}
&=-2\alpha \Phi_0
 \bigl[\cos(m\bar t+\Upsilon_{\rm asc})-\cos\Upsilon_{\rm asc}\bigr],
\\
\delta\eta
&=\frac{\alpha \Phi_0}{2}\biggl[
 \frac{\omega_b-2m}{\omega_b+m}\sin[(m+\omega_b)\bar t+\Upsilon_{\rm asc}]
 -\frac{\omega_b+2m}{\omega_b-m}\sin[(m-\omega_b)\bar t+\Upsilon_{\rm asc}]
+\frac{6m\omega_b}{\omega_b^2-m^2}\sin\Upsilon_{\rm asc}\biggr],
\\
\delta\kappa
&=\frac{\alpha \Phi_0}{2}\biggl[
 \frac{\omega_b-2m}{\omega_b+m}\cos[(m+\omega_b)\bar t+\Upsilon_{\rm asc}]
 +\frac{\omega_b+2m}{\omega_b-m}\cos[(m-\omega_b)\bar t+\Upsilon_{\rm asc}]
-\frac{2(\omega_b^2+2m^2)}{\omega_b^2-m^2}
 \cos\Upsilon_{\rm asc}\biggr],
\\
\delta T_{\rm asc}
&=-\frac{2\alpha \Phi_0}{m}
 \bigl[\sin(m\bar t+\Upsilon_{\rm asc})-\sin\Upsilon_{\rm asc}\bigr],
\\
\frac{\overline{\delta P_b}}{P_b}
&=\alpha \Phi_0\biggl[
 3\cos\Upsilon_{\rm asc}
 -\frac{7}{2m\bar t}
 \bigl[\sin(m\bar t+\Upsilon_{\rm asc})-\sin\Upsilon_{\rm asc}\bigr]\biggr].
\end{align}
\end{subequations}

\subsubsection{Eccentric orbits}
The five variations entering the eccentric timing template are
\begin{subequations}
\label{eq:scalar_linear_eccentric}
\begin{align}
\frac{\delta x}{x}
&=-2\alpha \Phi_0\bigl[\cos(m\tilde t+\Upsilon_0)-\cos\Upsilon_0\bigr]
-2\alpha \Phi_0\sum_{n=1}^{\infty}J_n
\bigl[(n\omega_b-2m)C_{n,1}^+
 +(n\omega_b+2m)C_{n,1}^-\bigr],\\
\delta e
&=-\frac{\alpha \Phi_0\sigma^2}{e}\sum_{n=1}^{\infty}J_n
\bigl[(n\omega_b-2m)C_{n,1}^+
 +(n\omega_b+2m)C_{n,1}^-\bigr],\\
\delta\omega
&=\frac{\alpha \Phi_0\sigma}{e}\sum_{n=1}^{\infty}J_n'
\bigl[(n\omega_b-2m)S_{n,1}^+
 +(n\omega_b+2m)S_{n,1}^-\bigr],\\
\delta T_0
&=-\frac{2\alpha \Phi_0}{m}
 \bigl[\sin(m\tilde t+\Upsilon_0)-\sin\Upsilon_0\bigr]
 -\frac{\alpha \Phi_0}{2\omega_b}\sum_{n=1}^{\infty}
 \bigl[(A_n+mB_n)S_{n,1}^+
 +(A_n-mB_n)S_{n,1}^-\bigr],\\
\frac{\overline{\delta P_b}}{P_b}
&=\alpha \Phi_0\biggl[
3\cos\Upsilon_0-\frac{7}{2m\tilde t}
\bigl[\sin(m\tilde t+\Upsilon_0)-\sin\Upsilon_0\bigr]\biggr]
\nonumber\\
&\quad-3\alpha \Phi_0\sum_{n=1}^{\infty}J_n
\biggl[\frac{n\omega_b-2m}{n\omega_b+m}
\Bigl(\cos\Upsilon_0-\frac{S_{n,1}^+}{\tilde t}\Bigr)
+\frac{n\omega_b+2m}{n\omega_b-m}
\Bigl(\cos\Upsilon_0-\frac{S_{n,1}^-}{\tilde t}\Bigr)\biggr].
\end{align}
\end{subequations}

\subsection{Quadratically coupled scalar ULDM}
\subsubsection{Circular orbits}

The five variations entering the circular timing template are
\begin{subequations}
\label{eq:scalar_quadratic_circular}
\begin{align}
\frac{\delta x}{x}
&=-\frac{\beta \Phi_0^2}{2}
 \bigl[\cos(2m\bar t+2\Upsilon_{\rm asc})-\cos(2\Upsilon_{\rm asc})\bigr],\\
\delta\eta
&=\frac{\beta \Phi_0^2}{2}\biggl\{
\frac12\sin(\omega_b\bar t)
+\frac{\omega_b/4-m}{\omega_b+2m}
\bigl[\sin[(2m+\omega_b)\bar t+2\Upsilon_{\rm asc}]
-\sin(2\Upsilon_{\rm asc})\bigr]
\nonumber\\
&\qquad+\frac{\omega_b/4+m}{\omega_b-2m}
\bigl[\sin(2\Upsilon_{\rm asc})
-\sin[(2m-\omega_b)\bar t+2\Upsilon_{\rm asc}]\bigr]\biggr\},\\
\delta\kappa
&=-\frac{\beta \Phi_0^2}{2}\biggl\{
\frac{1-\cos(\omega_b\bar t)}{2}
+\frac{\omega_b/4-m}{\omega_b+2m}
\bigl[\cos(2\Upsilon_{\rm asc})
-\cos[(2m+\omega_b)\bar t+2\Upsilon_{\rm asc}]\bigr]
\nonumber\\
&\qquad+\frac{\omega_b/4+m}{\omega_b-2m}
\bigl[\cos(2\Upsilon_{\rm asc})
-\cos[(2m-\omega_b)\bar t+2\Upsilon_{\rm asc}]\bigr]\biggr\},\\
\delta T_{\rm asc}
&=-\frac{\beta \Phi_0^2}{2}\biggl[
\bar t+\frac{\sin(2m\bar t+2\Upsilon_{\rm asc})
-\sin(2\Upsilon_{\rm asc})}{2m}\biggr],\\
\frac{\overline{\delta P_b}}{P_b}
&=\beta \Phi_0^2\biggl[
\frac34\cos(2\Upsilon_{\rm asc})-\frac18
-\frac{7}{16m\bar t}
\bigl[\sin(2m\bar t+2\Upsilon_{\rm asc})-\sin(2\Upsilon_{\rm asc})\bigr]\biggr].
\end{align}
\end{subequations}

\subsubsection{Eccentric orbits}

With the primitives of Eq. \eqref{eq:ecc_primitives} at $q=2$, the five variations entering the eccentric timing template are
\begin{subequations}
\label{eq:scalar_quadratic_eccentric}
\begin{align}
\frac{\delta x}{x}
&=-\frac{\beta \Phi_0^2}{2}
\bigl[\cos(2m\tilde t+2\Upsilon_0)-\cos(2\Upsilon_0)\bigr]
\nonumber\\
&\quad-2\beta \Phi_0^2\sum_{n=1}^{\infty}J_n\biggl[
\frac{1-\cos(n\ell)}{2}
+\Bigl(\frac{n\omega_b}{4}-m\Bigr)C_{n,2}^+
+\Bigl(\frac{n\omega_b}{4}+m\Bigr)C_{n,2}^-\biggr],\\
\delta e
&=-\frac{\beta \Phi_0^2\sigma^2}{e}
\sum_{n=1}^{\infty}J_n\biggl[
\frac{1-\cos(n\ell)}{2}
+\Bigl(\frac{n\omega_b}{4}-m\Bigr)C_{n,2}^+
+\Bigl(\frac{n\omega_b}{4}+m\Bigr)C_{n,2}^-\biggr],\\
\delta\omega
&=\frac{\beta \Phi_0^2\sigma}{e}
\sum_{n=1}^{\infty}J_n'\biggl[
\frac{\sin(n\ell)}{2}
+\Bigl(\frac{n\omega_b}{4}-m\Bigr)S_{n,2}^+
+\Bigl(\frac{n\omega_b}{4}+m\Bigr)S_{n,2}^-\biggr],\\
\delta T_0
&=-\frac{\beta \Phi_0^2}{2}
\biggl[\tilde t+\frac{\sin(2m\tilde t+2\Upsilon_0)
-\sin(2\Upsilon_0)}{2m}\biggr]
\nonumber\\
&\quad-\frac{\beta \Phi_0^2}{8\omega_b}\sum_{n=1}^{\infty}
\biggl[\frac{2A_n}{n\omega_b}\sin(n\ell)
+(A_n+2mB_n)S_{n,2}^+
+(A_n-2mB_n)S_{n,2}^-\biggr],\\
\frac{\overline{\delta P_b}}{P_b}
&=\beta \Phi_0^2\biggl[
\frac34\cos(2\Upsilon_0)-\frac18
-\frac{7}{16m\tilde t}
\bigl[\sin(2m\tilde t+2\Upsilon_0)-\sin(2\Upsilon_0)\bigr]\biggr]
\nonumber\\
&\quad-3\beta \Phi_0^2\sum_{n=1}^{\infty}J_n
\biggl[\frac12\Bigl(1-\frac{\sin(n\ell)}{n\ell}\Bigr)
+\sum_{\pm}\frac{n\omega_b/4\mp m}{n\omega_b\pm2m}
\Bigl(\cos(2\Upsilon_0)-\frac{S_{n,2}^\pm}{\tilde t}\Bigr)\biggr].
\end{align}
\end{subequations}

\subsection{Vector ULDM}

The scalar force of Eq. \eqref{eq:fuerza escalar} points along the relative velocity. Its components on the polar basis $(\hat r,\hat\theta,\hat z)$ therefore depend on the orbit only through the true anomaly, and $F_z=0$. The vector force points along $\hat n$, which is fixed in space while the polar basis turns with the pulsar. Its components in Eq. \eqref{eq:fuerza vectorial polar} carry $\theta$ through $\cos(\varphi-\theta)$ and $\sin(\varphi-\theta)$, and $F_z$ no longer vanishes.

\subsubsection{Circular orbits}

When $e\to0$ the periastron, and with it $\varphi$ and $\omega$ separately, is ill defined. Only the azimuth of $\hat n$ measured from the ascending node, $\varphi_{\rm asc}=\varphi+\omega$, survives. The five variations entering the circular timing template are
\begin{subequations}\label{eq:vector_circular}
\begin{align}
\frac{\delta x}{x}
&=-\frac{F_0}{a\omega_b}\sum_{\pm}\frac{1}{m\pm\omega_b}\biggl\{
\pm\sin\vartheta\bigl[\sin[(m\pm\omega_b)\bar t+\Upsilon_{\rm asc}\mp\varphi_{\rm asc}]
-\sin(\Upsilon_{\rm asc}\mp\varphi_{\rm asc})\bigr]
\nonumber\\
&\qquad+\frac{\cot\iota\cos\vartheta}{2}
\bigl[\cos\Upsilon_{\rm asc}-\cos[(m\pm\omega_b)\bar t+\Upsilon_{\rm asc}]\bigr]\biggr\},\\
\delta\eta
&=\frac{F_0\sin\vartheta}{a\omega_b}\biggl\{
\frac{3\cos\varphi_{\rm asc}}{2m}
\bigl[\cos\Upsilon_{\rm asc}
-\cos(m\bar t+\Upsilon_{\rm asc})\bigr]
\nonumber\\
&\qquad-\frac14\sum_{\pm}
\frac{\cos(\Upsilon_{\rm asc}\mp\varphi_{\rm asc})
-\cos[(m\pm2\omega_b)\bar t+\Upsilon_{\rm asc}\mp\varphi_{\rm asc}]}{m\pm2\omega_b}\biggr\},\\
\delta\kappa
&=-\frac{F_0\sin\vartheta}{a\omega_b}\biggl\{
\frac{3\sin\varphi_{\rm asc}}{2m}
\bigl[\cos\Upsilon_{\rm asc}
-\cos(m\bar t+\Upsilon_{\rm asc})\bigr]
\nonumber\\
&\qquad+\frac14\sum_{\pm}\pm
\frac{\sin[(m\pm2\omega_b)\bar t+\Upsilon_{\rm asc}\mp\varphi_{\rm asc}]
-\sin(\Upsilon_{\rm asc}\mp\varphi_{\rm asc})}{m\pm2\omega_b}\biggr\},\\
\delta T_{\rm asc}
&=-\frac{F_0}{a\omega_b^2}\sum_{\pm}\frac{1}{m\pm\omega_b}\biggl\{
\mp\frac{\cot\iota\cos\vartheta}{2}
\bigl[\sin[(m\pm\omega_b)\bar t+\Upsilon_{\rm asc}]-\sin\Upsilon_{\rm asc}\bigr]
\nonumber\\
&\qquad+\sin\vartheta\bigl[\cos(\Upsilon_{\rm asc}\mp\varphi_{\rm asc})
-\cos[(m\pm\omega_b)\bar t+\Upsilon_{\rm asc}\mp\varphi_{\rm asc}]\bigr]\biggr\},\\
\frac{\overline{\delta P_b}}{P_b}
&=-\frac{3F_0\sin\vartheta}{2a\omega_b}\sum_{\pm}\frac{\pm1}{m\pm\omega_b}
\biggl\{\frac{\cos(\Upsilon_{\rm asc}\mp\varphi_{\rm asc})
-\cos[(m\pm\omega_b)\bar t+\Upsilon_{\rm asc}\mp\varphi_{\rm asc}]}{(m\pm\omega_b)\bar t}
-\sin(\Upsilon_{\rm asc}\mp\varphi_{\rm asc})\biggr\}.
\end{align}
\end{subequations}
The terms with $m-\omega_b$ resonate in $x$, $T_{\rm asc}$ and $P_b$, and those with $m-2\omega_b$ in $\eta$ and $\kappa$.

\subsubsection{Eccentric orbits}

For $e\neq0$ the angles $\varphi$ and $\omega$ are well defined separately, and we work with the polar components of Eq. \eqref{eq:fuerza vectorial polar}. Expanding $\cos(\varphi-\theta)=\cos\varphi\cos\theta+\sin\varphi\sin\theta$ and $\sin(\varphi-\theta)=\sin\varphi\cos\theta-\cos\varphi\sin\theta$, and inserting the components into Eqs. \eqref{eq:gauss_equations}, each rate becomes $\sin(m\tilde t+\Upsilon_0)$ times a function of the orbit. Using $1+e\cos\theta=\sigma^2a/r$ and $\cos E=(e+\cos\theta)/(1+e\cos\theta)$ to simplify,
\begin{subequations}\label{eq:vector_ecc_rates}
\begin{align}
\frac{\dot a}{a}
&=-\frac{2F_0\sin\vartheta}{a\omega_b\sigma}\,\sin(m\tilde t+\Upsilon_0)
\bigl[-\cos\varphi\,\sin\theta+\sin\varphi\,(e+\cos\theta)\bigr],\\
\dot e
&=-\frac{\sigma F_0\sin\vartheta}{a\omega_b}\,\sin(m\tilde t+\Upsilon_0)
\bigl[-\cos\varphi\,\sin\theta\cos E+\sin\varphi\,(1+\cos\theta\cos E)\bigr],\\
\dot\omega
&=\frac{\sigma F_0\sin\vartheta}{ae\omega_b}\,\sin(m\tilde t+\Upsilon_0)
\biggl[\cos\varphi\Bigl(1+\frac{\sin^2\theta}{1+e\cos\theta}\Bigr)
-\sin\varphi\,\frac{\sin\theta\cos\theta}{1+e\cos\theta}\biggr]-\cos\iota\,\dot\Omega,\\
\dot\iota
&=-\frac{F_0\cos\vartheta}{a\omega_b\sigma}\,\sin(m\tilde t+\Upsilon_0)\,
\frac ra\bigl[\cos\omega\cos\theta-\sin\omega\sin\theta\bigr],\\
\dot\Omega
&=-\frac{F_0\cos\vartheta}{a\omega_b\sigma\sin\iota}\,\sin(m\tilde t+\Upsilon_0)\,
\frac ra\bigl[\sin\omega\cos\theta+\cos\omega\sin\theta\bigr].
\end{align}
\end{subequations}
The timing model fits $x=a_1\sin\iota$, whose rate is
\begin{equation}
\frac{\dot x}{x}=\frac{\dot a}{a}+\cot\iota\,\dot\iota .
\end{equation}
For the time of periastron passage, Eq. \eqref{eq:gauss_epsilon} and $\dot\varpi=\dot\omega+\dot\Omega$ give
\begin{equation}
\begin{aligned}
\dot\epsilon_1-\dot\varpi
&=-\frac{2r}{a^2\omega_b}F_r-\sigma\bigl(\dot\omega+\cos\iota\,\dot\Omega\bigr)\\
&=\frac{2F_0\sin\vartheta}{a\omega_b}\,\sin(m\tilde t+\Upsilon_0)\,
\frac ra\bigl[\cos\varphi\cos\theta+\sin\varphi\sin\theta\bigr]\\
&\quad-\frac{\sigma^2F_0\sin\vartheta}{ae\omega_b}\,\sin(m\tilde t+\Upsilon_0)
\biggl[\cos\varphi\Bigl(1+\frac{\sin^2\theta}{1+e\cos\theta}\Bigr)
-\sin\varphi\,\frac{\sin\theta\cos\theta}{1+e\cos\theta}\biggr],
\end{aligned}
\end{equation}
in which the $\dot\Omega$ terms have canceled, so that $\delta T_0$ does not depend on the inclination.

Besides the functions expanded in Eq. \eqref{eq:bessel_series}, the brackets contain four combinations of the anomalies. The Kepler relations
\begin{equation*}
\cos E=e+\frac ra\cos\theta,\qquad 1+e\cos\theta=\sigma^2\,\frac ar
\end{equation*}
reduce them to two new products,
\begin{align*}
\sin\theta\cos E&=e\sin\theta+\frac ra\sin\theta\cos\theta,
&
1+\cos\theta\cos E&=1+e\cos\theta+\frac ra-\frac ra\sin^2\theta,\\
\frac{\sin^2\theta}{1+e\cos\theta}&=\frac{1}{\sigma^2}\,\frac ra\sin^2\theta,
&
\frac{\sin\theta\cos\theta}{1+e\cos\theta}&=\frac{1}{\sigma^2}\,\frac ra\sin\theta\cos\theta,
\end{align*}
whose expansions are
\begin{subequations}\label{eq:series_vector}
\begin{align}
\frac ra\sin^2\theta
&=\frac{\sigma^2}{2}-\frac{2\sigma^2}{e^2}\sum_{n=1}^{\infty}\Bigl(\sigma^2J_n-\frac enJ_n'\Bigr)\cos(n\ell),\\
\frac ra\sin\theta\cos\theta
&=-\frac{2\sigma}{e^2}\sum_{n=1}^{\infty}\Bigl(\frac{J_n}{n}-e\sigma^2J_n'\Bigr)\sin(n\ell).
\end{align}
\end{subequations}
In terms of the eccentric anomaly both products carry a factor $a/r$,
\begin{equation*}
\frac ra\sin^2\theta=\sigma^2\,\frac ar\,\sin^2E,\qquad
\frac ra\sin\theta\cos\theta=\sigma\,\frac ar\,\sin E\,(\cos E-e),
\end{equation*}
which cancels in their Fourier integrals against $d\ell=(r/a)\,dE$. The coefficients then reduce to Bessel's integral,
\begin{equation*}
J_k(ne)=\frac1\pi\int_0^\pi\cos\bigl(kE-ne\sin E\bigr)\,dE ,
\end{equation*}
with $k=n$, $n\pm1$ and $n\pm2$, and the recurrence relations express $J_{n\pm1}$ and $J_{n\pm2}$ through $J_n$ and $J_n'$.

Every rate is then $\sin(m\tilde t+\Upsilon_0)$ times a constant, $\cos(n\ell)$ or $\sin(n\ell)$. Integrated from periastron passage, the harmonics give the primitives of Eq. \eqref{eq:ecc_primitives} at $q=1$,
\begin{equation}\label{eq:vector_ecc_integrals}
\int_0^{\tilde t}\sin(m\tilde t'+\Upsilon_0)\cos(n\ell')\,d\tilde t'=\frac12\bigl(C_{n,1}^+-C_{n,1}^-\bigr),
\qquad
\int_0^{\tilde t}\sin(m\tilde t'+\Upsilon_0)\sin(n\ell')\,d\tilde t'=-\frac12\bigl(S_{n,1}^+-S_{n,1}^-\bigr),
\end{equation}
which resonate at $m=n\omega_b$. The constant terms give $[\cos\Upsilon_0-\cos(m\tilde t+\Upsilon_0)]/m$, which never resonates. The five variations entering the eccentric timing template are
\begin{subequations}\label{eq:vector_eccentric}
\begin{align}
\frac{\delta x}{x}
&=-\frac{2F_0\sin\vartheta}{a\omega_b}\sum_{n=1}^{\infty}\sum_\pm\pm
\Bigl[J_n'\cos\varphi\,S_{n,1}^\pm+\frac{\sigma}{e}J_n\sin\varphi\,C_{n,1}^\pm\Bigr]
\nonumber\\
&\quad+\frac{F_0\cot\iota\cos\vartheta}{a\omega_b\sigma}\biggl\{
\frac{3e\cos\omega}{2m}\bigl[\cos\Upsilon_0-\cos(m\tilde t+\Upsilon_0)\bigr]
-\sum_{n=1}^{\infty}\sum_\pm\pm\Bigl[\frac{J_n'}{n}\cos\omega\,C_{n,1}^\pm+\frac{\sigma J_n}{ne}\sin\omega\,S_{n,1}^\pm\Bigr]\biggr\},\\
\delta e
&=-\frac{\sigma F_0\sin\vartheta}{a\omega_b}\biggl\{
\frac{3\sin\varphi}{2m}\bigl[\cos\Upsilon_0-\cos(m\tilde t+\Upsilon_0)\bigr]
\nonumber\\
&\qquad+\frac{1}{e^2}\sum_{n=1}^{\infty}\sum_\pm\pm
\Bigl[\Bigl(\sigma^2J_n-\frac enJ_n'\Bigr)\sin\varphi\,C_{n,1}^\pm
-\sigma\Bigl(\frac{J_n}{n}-eJ_n'\Bigr)\cos\varphi\,S_{n,1}^\pm\Bigr]\biggr\},\\
\delta\omega
&=\frac{\sigma F_0\sin\vartheta}{ae\omega_b}\biggl\{
\frac{3\cos\varphi}{2m}\bigl[\cos\Upsilon_0-\cos(m\tilde t+\Upsilon_0)\bigr]
\nonumber\\
&\qquad-\frac{1}{e^2}\sum_{n=1}^{\infty}\sum_\pm\pm
\Bigl[\Bigl(\sigma^2J_n-\frac enJ_n'\Bigr)\cos\varphi\,C_{n,1}^\pm
+\frac1\sigma\Bigl(\frac{J_n}{n}-e\sigma^2J_n'\Bigr)\sin\varphi\,S_{n,1}^\pm\Bigr]\biggr\}
\nonumber\\
&\quad-\frac{F_0\cot\iota\cos\vartheta}{a\omega_b\sigma}\biggl\{
\frac{3e\sin\omega}{2m}\bigl[\cos\Upsilon_0-\cos(m\tilde t+\Upsilon_0)\bigr]
-\sum_{n=1}^{\infty}\sum_\pm\pm\Bigl[\frac{J_n'}{n}\sin\omega\,C_{n,1}^\pm-\frac{\sigma J_n}{ne}\cos\omega\,S_{n,1}^\pm\Bigr]\biggr\},\\
\delta T_0
&=\frac{F_0\sin\vartheta}{a\omega_b^2}\biggl\{
\frac{3(1+e^2)\cos\varphi}{2em}\bigl[\cos\Upsilon_0-\cos(m\tilde t+\Upsilon_0)\bigr]
\nonumber\\
&\qquad-\sum_{n=1}^{\infty}\sum_\pm\pm
\Bigl[\Bigl(\frac{\sigma^4}{e^3}J_n+\frac{3e^2-1}{ne^2}J_n'\Bigr)\cos\varphi\,C_{n,1}^\pm
-\frac{\sigma}{e^3}\Bigl(\frac{2e^2-1}{n}J_n+e\sigma^2J_n'\Bigr)\sin\varphi\,S_{n,1}^\pm\Bigr]\biggr\},\\
\frac{\overline{\delta P_b}}{P_b}
&=-\frac{3F_0\sin\vartheta}{a\omega_b}\sum_{n=1}^{\infty}\sum_\pm\frac{\pm1}{n\omega_b\pm m}
\Bigl[J_n'\cos\varphi\Bigl(\frac{C_{n,1}^\pm}{\tilde t}\mp\sin\Upsilon_0\Bigr)
+\frac{\sigma}{e}J_n\sin\varphi\Bigl(\cos\Upsilon_0-\frac{S_{n,1}^\pm}{\tilde t}\Bigr)\Bigr].
\end{align}
\end{subequations}
\flushbottom
\twocolumngrid
\section{The sample}
\label{app:sample}

\begin{table*}[p]
\centering
\footnotesize
\setlength{\tabcolsep}{2.9pt}
\caption{The $50$ binary pulsars of the sample. Model is the binary model of the published solution; $P_b$ the orbital period; $x$ the projected semi-major axis in light-seconds; $e$ the eccentricity; $N_{\rm TOA}$ the number of wideband TOAs; $T_{\mathrm{obs}}$ the time span covered by them; $m_{\rm res}=\omega_b$ the fundamental resonance; Comp.\ the companion type from the ATNF catalogue \cite{ATNF2025}, with a dash where the catalogue gives none; $\mathrm{M}_1$ and $\mathrm{M}_2$ the pulsar and companion masses; $\iota$ the orbital inclination, folded into $[0^\circ,90^\circ]$. The last column gives the source of the masses and inclination: a reference to the literature; NG for the NANOGrav 15 yr solution \cite{NANOGrav15data}, through its Shapiro delay (or, for PSR J1713$+$0747, its kinematic change of $x$); or a dash for the ATNF convention of case (C). A dagger marks assumed values, $\mathrm{M}_1=1.35\,M_\odot$ and $\iota=60^\circ$; an asterisk marks values derived from the mass function, Eq. \eqref{eq:massfunction}, given the other two. A double dagger after the PSR name marks the nine solutions with a fitted $\dot P_b$, held at its published value in our analysis.}
\label{tab:sample}
\begin{tabular}{lcrrrrrrcrrrc}
\toprule
PSR & Model & $P_b$ [d] & $x$ [lt-s] & $e$ & $N_{\rm TOA}$ & $T_{\mathrm{obs}}$ [yr] & $m_{\rm res}$ [eV] & Comp. & $\mathrm{M}_1$ [$M_\odot$] & $\mathrm{M}_2$ [$M_\odot$] & $\iota$ [$^\circ$] & Ref. \\
\midrule
B1855$+$09 & ELL1 & 12.33 & 9.231 & $2.2\times10^{-5}$ & 364 & 15.6 & $3.9\times10^{-21}$ & He & 1.37 & 0.244$^{\ast}$ & 87.3 & \cite{Arzoumanian2018}, NG \\
B1953$+$29 & DD & 117.3 & 31.41 & $3.3\times10^{-4}$ & 173 & 9.0 & $4.1\times10^{-22}$ & He & 1.35$^{\dagger}$ & 0.208$^{\ast}$ & 60$^{\dagger}$ & -- \\
J0023$+$0923 & ELL1 & 0.1388 & 0.03484 & $8.4\times10^{-6}$ & 822 & 8.9 & $3.4\times10^{-19}$ & UL & 1.35$^{\dagger}$ & 0.019$^{\ast}$ & 60$^{\dagger}$ & -- \\
J0406$+$3039 & ELL1 & 6.956 & 2.319 & $1.2\times10^{-5}$ & 71 & 3.6 & $6.9\times10^{-21}$ & -- & 1.35$^{\dagger}$ & 0.096$^{\ast}$ & 60$^{\dagger}$ & -- \\
J0437$-$4715 & DD & 5.741 & 3.367 & $1.9\times10^{-5}$ & 117 & 4.8 & $8.3\times10^{-21}$ & He & 1.418 & 0.221 & 42.6$^{\ast}$ & \cite{Reardon2024} \\
J0509$+$0856 & ELL1 & 4.908 & 2.458 & $2.4\times10^{-5}$ & 66 & 3.6 & $9.8\times10^{-21}$ & He & 1.35$^{\dagger}$ & 0.131$^{\ast}$ & 60$^{\dagger}$ & -- \\
J0557$+$1551 & ELL1 & 4.847 & 4.054 & $9.4\times10^{-6}$ & 47 & 4.6 & $9.9\times10^{-21}$ & -- & 1.35$^{\dagger}$ & 0.227$^{\ast}$ & 60$^{\dagger}$ & -- \\
J0605$+$3757 & DD & 55.67 & 18.95 & $3.8\times10^{-4}$ & 47 & 3.4 & $8.6\times10^{-22}$ & He & 1.35$^{\dagger}$ & 0.206$^{\ast}$ & 60$^{\dagger}$ & -- \\
J0610$-$2100 & ELL1 & 0.2860 & 0.07349 & $1.6\times10^{-5}$ & 214 & 3.4 & $1.7\times10^{-19}$ & UL & 1.35$^{\dagger}$ & 0.022$^{\ast}$ & 72.9 & \cite{vanderWateren2022} \\
J0613$-$0200$^{\ddagger}$ & ELL1 & 1.199 & 1.091 & $4.2\times10^{-6}$ & 423 & 15.0 & $4.0\times10^{-20}$ & He & 1.35$^{\dagger}$ & 0.144$^{\ast}$ & 64.4 & NG \\
J0614$-$3329 & DD & 53.58 & 27.64 & $1.8\times10^{-4}$ & 56 & 2.4 & $8.9\times10^{-22}$ & He & 1.35$^{\dagger}$ & 0.275$^{\ast}$ & 90.0 & \cite{Bassa2016} \\
J0636$+$5128 & ELL1 & 0.06655 & 0.008986 & $1.3\times10^{-5}$ & 1221 & 6.3 & $7.2\times10^{-19}$ & UL & 1.35$^{\dagger}$ & $7.9\times10^{-3}$$^{\ast}$ & 60$^{\dagger}$ & -- \\
J0709$+$0458 & DD & 4.367 & 15.72 & $2.3\times10^{-4}$ & 91 & 4.6 & $1.1\times10^{-20}$ & CO & 1.35$^{\dagger}$ & 1.165$^{\ast}$ & 73.0 & \cite{Martinez2019} \\
J0740$+$6620$^{\ddagger}$ & ELL1 & 4.767 & 3.978 & $5.9\times10^{-6}$ & 360 & 6.3 & $1.0\times10^{-20}$ & He & 2.08 & 0.253$^{\ast}$ & 87.7 & \cite{Fonseca2021}, NG \\
J1012$+$5307$^{\ddagger}$ & ELL1 & 0.6047 & 0.5818 & $1.4\times10^{-6}$ & 624 & 15.5 & $7.9\times10^{-20}$ & He & 1.72 & 0.165 & 50.4$^{\ast}$ & \cite{MataSanchez2020} \\
J1012$-$4235 & DD & 37.97 & 21.26 & $3.5\times10^{-4}$ & 65 & 3.4 & $1.3\times10^{-21}$ & He & 1.44 & 0.277$^{\ast}$ & 87.3 & \cite{Gautam2024}, NG \\
J1022$+$1001 & DD & 7.805 & 16.77 & $9.7\times10^{-5}$ & 116 & 5.6 & $6.1\times10^{-21}$ & CO & 1.35$^{\dagger}$ & 1.172$^{\ast}$ & 43.6 & \cite{Reardon2016} \\
J1125$+$7819 & ELL1 & 15.36 & 12.19 & $1.3\times10^{-5}$ & 206 & 6.3 & $3.1\times10^{-21}$ & He & 1.35$^{\dagger}$ & 0.330$^{\ast}$ & 60$^{\dagger}$ & -- \\
J1312$+$0051 & BT & 38.50 & 14.75 & $2.4\times10^{-4}$ & 76 & 4.6 & $1.2\times10^{-21}$ & He & 1.35$^{\dagger}$ & 0.205$^{\ast}$ & 60$^{\dagger}$ & -- \\
J1455$-$3330 & DD & 76.17 & 32.36 & $1.7\times10^{-4}$ & 357 & 15.7 & $6.3\times10^{-22}$ & He & 1.39 & 0.293 & 62.9$^{\ast}$ & \cite{Pillay2026} \\
J1600$-$3053 & DD & 14.35 & 8.802 & $1.7\times10^{-4}$ & 481 & 12.5 & $3.3\times10^{-21}$ & He & 1.35$^{\dagger}$ & 0.223$^{\ast}$ & 68.0 & NG \\
J1614$-$2230$^{\ddagger}$ & ELL1 & 8.687 & 11.29 & $1.3\times10^{-6}$ & 367 & 11.5 & $5.5\times10^{-21}$ & CO & 1.94 & 0.495$^{\ast}$ & 89.2 & \cite{Shamohammadi2023}, NG \\
J1630$+$3734 & DD & 12.53 & 9.039 & $5.2\times10^{-5}$ & 71 & 3.5 & $3.8\times10^{-21}$ & He & 1.35$^{\dagger}$ & 0.241$^{\ast}$ & 76.4 & NG \\
J1640$+$2224 & DD & 175.5 & 55.33 & $8.0\times10^{-4}$ & 609 & 15.5 & $2.7\times10^{-22}$ & He & 1.35$^{\dagger}$ & 0.276$^{\ast}$ & 65.0 & NG \\
J1643$-$1224 & DD & 147.0 & 25.07 & $5.1\times10^{-4}$ & 478 & 15.7 & $3.3\times10^{-22}$ & He & 1.35$^{\dagger}$ & 0.397$^{\ast}$ & 19.7 & \cite{Reardon2016} \\
J1705$-$1903 & ELL1 & 0.1840 & 0.1044 & $4.3\times10^{-6}$ & 253 & 3.7 & $2.6\times10^{-19}$ & -- & 1.35$^{\dagger}$ & 0.048$^{\ast}$ & 60$^{\dagger}$ & -- \\
J1713$+$0747 & DDK & 67.83 & 32.34 & $7.5\times10^{-5}$ & 1493 & 15.5 & $7.1\times10^{-22}$ & He & 1.45$^{\ast}$ & 0.307 & 70.7 & NG \\
J1719$-$1438$^{\ddagger}$ & ELL1 & 0.09071 & 0.001821 & $2.8\times10^{-4}$ & 463 & 3.4 & $5.3\times10^{-19}$ & UL & 1.35$^{\dagger}$ & $1.3\times10^{-3}$$^{\ast}$ & 60$^{\dagger}$ & -- \\
J1738$+$0333 & ELL1 & 0.3548 & 0.3434 & $1.5\times10^{-6}$ & 336 & 10.7 & $1.3\times10^{-19}$ & He & 1.47 & 0.181 & 32.8$^{\ast}$ & \cite{Antoniadis2012} \\
J1741$+$1351$^{\ddagger}$ & ELL1 & 16.34 & 11.00 & $1.0\times10^{-5}$ & 208 & 9.1 & $2.9\times10^{-21}$ & He & 1.35$^{\dagger}$ & 0.246$^{\ast}$ & 76.4 & NG \\
J1745$+$1017 & ELL1 & 0.7302 & 0.08817 & $2.5\times10^{-5}$ & 91 & 4.5 & $6.6\times10^{-20}$ & UL & 1.35$^{\dagger}$ & 0.016$^{\ast}$ & 60$^{\dagger}$ & -- \\
J1751$-$2857 & DD & 110.7 & 32.53 & $1.3\times10^{-4}$ & 89 & 3.5 & $4.3\times10^{-22}$ & He & 1.35$^{\dagger}$ & 0.226$^{\ast}$ & 60$^{\dagger}$ & -- \\
J1802$-$2124 & ELL1H & 0.6989 & 3.719 & $3.3\times10^{-6}$ & 126 & 3.5 & $6.8\times10^{-20}$ & CO & 1.24 & 0.780 & 82.1$^{\ast}$ & \cite{Ferdman2010} \\
J1811$-$2405 & ELL1 & 6.272 & 5.706 & $1.2\times10^{-6}$ & 103 & 3.5 & $7.6\times10^{-21}$ & He & 1.35$^{\dagger}$ & 0.241$^{\ast}$ & 76.4 & NG \\
J1853$+$1303 & DD & 115.7 & 40.77 & $2.4\times10^{-5}$ & 184 & 9.0 & $4.1\times10^{-22}$ & He & 1.35$^{\dagger}$ & 0.281$^{\ast}$ & 60$^{\dagger}$ & -- \\
J1903$+$0327 & DD & 95.17 & 105.6 & $4.4\times10^{-1}$ & 226 & 10.7 & $5.0\times10^{-22}$ & MS & 1.667 & 1.026$^{\ast}$ & 78.1 & \cite{Freire2011}, NG \\
J1909$-$3744$^{\ddagger}$ & ELL1 & 1.533 & 1.898 & $1.1\times10^{-7}$ & 833 & 15.6 & $3.1\times10^{-20}$ & He & 1.492 & 0.209$^{\ast}$ & 86.4 & \cite{Liu2020}, NG \\
J1910$+$1256 & DD & 58.47 & 21.13 & $2.3\times10^{-4}$ & 216 & 11.4 & $8.2\times10^{-22}$ & He & 1.35$^{\dagger}$ & 0.287$^{\ast}$ & 44.0 & \cite{Gonzalez2011} \\
J1918$-$0642 & DD & 10.91 & 8.350 & $2.0\times10^{-5}$ & 487 & 15.5 & $4.4\times10^{-21}$ & He & 1.29 & 0.231$^{\ast}$ & 84.7 & \cite{Arzoumanian2018}, NG \\
J1946$+$3417 & DD & 27.02 & 13.87 & $1.3\times10^{-1}$ & 129 & 5.7 & $1.8\times10^{-21}$ & He & 1.828 & 0.273$^{\ast}$ & 71.2 & \cite{Barr2017}, NG \\
J2017$+$0603 & ELL1 & 2.198 & 2.193 & $7.0\times10^{-6}$ & 154 & 8.3 & $2.2\times10^{-20}$ & He & 1.35$^{\dagger}$ & 0.182$^{\ast}$ & 76.1 & NG \\
J2033$+$1734 & DD & 56.31 & 20.16 & $1.3\times10^{-4}$ & 133 & 7.0 & $8.5\times10^{-22}$ & He & 1.35$^{\dagger}$ & 0.219$^{\ast}$ & 60$^{\dagger}$ & -- \\
J2043$+$1711$^{\ddagger}$ & ELL1 & 1.482 & 1.624 & $5.0\times10^{-6}$ & 459 & 9.1 & $3.2\times10^{-20}$ & He & 1.6 & 0.190$^{\ast}$ & 82.0 & \cite{Kaiser2026}, NG \\
J2145$-$0750 & ELL1H & 6.839 & 10.16 & $1.9\times10^{-5}$ & 400 & 15.5 & $7.0\times10^{-21}$ & CO & 1.3 & 0.830 & 35.2$^{\ast}$ & \cite{Fonseca2016} \\
J2214$+$3000 & ELL1 & 0.4166 & 0.05908 & $8.3\times10^{-6}$ & 293 & 8.6 & $1.1\times10^{-19}$ & UL & 1.35$^{\dagger}$ & 0.015$^{\ast}$ & 60$^{\dagger}$ & -- \\
J2229$+$2643 & DD & 93.02 & 18.91 & $2.6\times10^{-4}$ & 151 & 7.0 & $5.1\times10^{-22}$ & He & 1.35$^{\dagger}$ & 0.142$^{\ast}$ & 60$^{\dagger}$ & -- \\
J2234$+$0611$^{\ddagger}$ & DD & 32.00 & 13.94 & $1.3\times10^{-1}$ & 133 & 6.5 & $1.5\times10^{-21}$ & He & 1.353 & 0.298 & 41.6$^{\ast}$ & \cite{Stovall2019} \\
J2234$+$0944 & ELL1 & 0.4197 & 0.06843 & $4.5\times10^{-6}$ & 245 & 7.1 & $1.1\times10^{-19}$ & UL & 1.35$^{\dagger}$ & 0.018$^{\ast}$ & 60$^{\dagger}$ & -- \\
J2302$+$4442 & DD & 125.9 & 51.43 & $5.0\times10^{-4}$ & 236 & 8.1 & $3.8\times10^{-22}$ & He & 1.35$^{\dagger}$ & 0.297$^{\ast}$ & 79.6 & NG \\
J2317$+$1439 & ELL1H & 2.459 & 2.314 & $4.4\times10^{-7}$ & 708 & 15.6 & $1.9\times10^{-20}$ & He & 1.35$^{\dagger}$ & 0.201$^{\ast}$ & 60$^{\dagger}$ & -- \\
\bottomrule
\end{tabular}
\end{table*}

Table \ref{tab:sample} lists the $50$ binaries of the NANOGrav 15 yr wideband data set that enter the analysis, together with the masses and inclination used by the vector template. The orbital parameters are those of the published timing solutions \cite{NANOGrav15data}. The eccentricity is $\sqrt{\eta^2+\kappa^2}$ for systems published in the ELL1 family. The column $m_{\rm res}$ gives the fundamental resonance $m=\omega_b$; for the quadratic coupling the corresponding mass is $\omega_b/2$. The companion type is the classification of the ATNF catalogue \cite{ATNF2025}: helium (He) or carbon--oxygen (CO) white dwarf, main-sequence star (MS), or ultralight companion (UL), the latter comprising the black-widow systems and the planetary companion of PSR J1719$-$1438.

The scalar templates depend on the orbit only through $\omega_b$, $e$, $\omega$ and the reference epochs. The vector template also needs the semi-major axis $a$ of the relative orbit and the inclination $\iota$, which enter through $F_0/a$ and $\cot\iota$ (Appendix \ref{app_orbital}). The semi-major axis follows from Kepler's third law,
\begin{equation}
    a^3=\frac{G\,\mathrm{M}_{\rm T}}{\omega_b^2},\qquad \mathrm{M}_{\rm T}=\mathrm{M}_1+\mathrm{M}_2,
\end{equation}
so it requires the total mass of the binary. The three quantities $\mathrm{M}_1$, $\mathrm{M}_2$ and $\iota$ are not independent. The Keplerian parameters of any timing solution fix the mass function
\begin{equation}
    f \equiv \frac{(\mathrm{M}_2\sin\iota)^3}{(\mathrm{M}_1+\mathrm{M}_2)^2} = \frac{4\pi^2}{T_\odot}\,\frac{x^3}{P_b^2},
    \label{eq:massfunction}
\end{equation}
with $T_\odot\equiv G M_\odot/c^3=4.925490947\,\mu$s, the masses in solar units, $x$ in light-seconds and $P_b$ in seconds. Because $x$ and $P_b$ are measured to parts in $10^{5}$ or better, $f$ is known with high precision, and any two of the three quantities determine the third. Setting $\sin\iota=1$ in Eq. \eqref{eq:massfunction} gives the minimum companion mass compatible with a given $\mathrm{M}_1$. For every system we therefore fix the two quantities that are best known and derive the third from the mass function, in one of three cases:
\begin{enumerate}
    \item[(A)] Both masses are published, and the wideband timing solution does not measure $\sin\iota$ (7 systems). The masses come from a single work that measures them together, through the Shapiro delay, the relativistic periastron advance, or optical spectroscopy of the companion combined with pulsar timing. We take $\mathrm{M}_1$ and $\mathrm{M}_2$ from that work and obtain $\iota$ from Eq. \eqref{eq:massfunction}.
    \item[(B)] The inclination is measured or constrained (24 systems). For these systems we fix $\iota$, take $\mathrm{M}_1$ from the literature where it is published and $\mathrm{M}_1=1.35\,M_\odot$ otherwise, and solve Eq. \eqref{eq:massfunction} for $\mathrm{M}_2$. The inclination comes from one of five sources:
    \begin{enumerate}
        \item[(i)] \emph{Shapiro delay in the wideband timing solution (17 systems).} The shape of the Shapiro delay gives $\sin\iota$ and its amplitude gives $\mathrm{M}_2$. The amplitude is much less precise, with uncertainties on $\mathrm{M}_2$ of up to $46\%$, so we keep only $\sin\iota$. We use it only if the fit is physically consistent: within $3\sigma$, the fitted $\mathrm{M}_2$ and $\sin\iota$ must give, through Eq. \eqref{eq:massfunction}, a pulsar mass between $1.0$ and $2.5\,M_\odot$. PSR J1853$+$1303 fails this test. Its fitted $\mathrm{M}_2=0.019\pm0.014\,M_\odot$ would require $\mathrm{M}_1\leq0.13\,M_\odot$, so we do not regard its Shapiro delay as detected, discard its $\sin\iota$ as well, and treat it as case (C).
        \item[(ii)] \emph{Kinematic change of $x$ (2 systems).} The proper motion of the binary slowly changes the orientation of the orbit relative to the line of sight. This changes the projected semi-major axis $x$ at a rate that depends on $\iota$ \cite{Kopeikin1996}. For PSR J1910$+$1256 we take the resulting inclination from Ref. \cite{Gonzalez2011}. For PSR J1713$+$0747, the only DDK solution in the sample, the timing solution gives both $\iota$ and $\mathrm{M}_2$, so we take both and solve Eq. \eqref{eq:massfunction} for $\mathrm{M}_1$.
        \item[(iii)] \emph{Inclination from the literature (2 systems).} PSRs J0709$+$0458 \cite{Martinez2019} and J1022$+$1001 \cite{Reardon2016}.
        \item[(iv)] \emph{Spectroscopy of the companion (1 system).} For PSR J0614$-$3329, the spectroscopic mass of the white dwarf is $0.24\pm0.04\,M_\odot$ \cite{Bassa2016}. With $\mathrm{M}_1=1.35\,M_\odot$, Eq. \eqref{eq:massfunction} gives its smallest companion mass at $\iota=90^\circ$, and even this value is larger than the spectroscopic one. Any other inclination would make the disagreement larger, so we adopt $\iota=90^\circ$.
        \item[(v)] \emph{Bound on the inclination only (2 systems).} PSR J1643$-$1224 has $\iota\leq28^\circ$ \cite{Reardon2016} and PSR J0610$-$2100 has $\iota>54^\circ$ \cite{vanderWateren2022}. We assume isotropic orbital orientations, for which $\cos\iota$ is uniform, and take the median inclination within the allowed range. This results in $\iota=19.7^\circ$ and $\iota=72.9^\circ$, respectively.
    \end{enumerate}
    
    \item[(C)] Neither the masses nor the inclination are measured (19 systems). We adopt the convention of the ATNF catalogue \cite{ATNF2025}: $\mathrm{M}_1=1.35\,M_\odot$ and $\iota=60^\circ$, the median of an isotropic distribution of orbital orientations. Equation \eqref{eq:massfunction} then gives the median companion mass.
\end{enumerate}

In every case the adopted triplet satisfies Eq. \eqref{eq:massfunction}. The small companion masses of the ultralight systems, between $1.3\times10^{-3}$ and $2.2\times10^{-2}\,M_\odot$, are not an artifact of this procedure: they follow from their small mass functions and correspond to the black-widow and planetary companions of these pulsars.

\begin{figure}[]
    \centering
    \includegraphics[width=\linewidth]{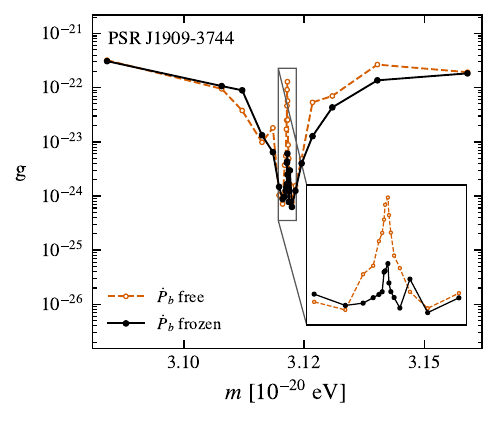}
    \caption{Combined $95\%$ credible limit on $g$ around the $m=\omega_b$ resonance of PSR J1909$-$3744, with $\dot P_b$ of the nine binaries whose timing solutions fit it held at the published value (black) and marginalized (orange, dashed).}
    \label{fig:pbdot}
\end{figure}

\begin{figure*}[]
    \centering
    \includegraphics[width=0.9\linewidth]{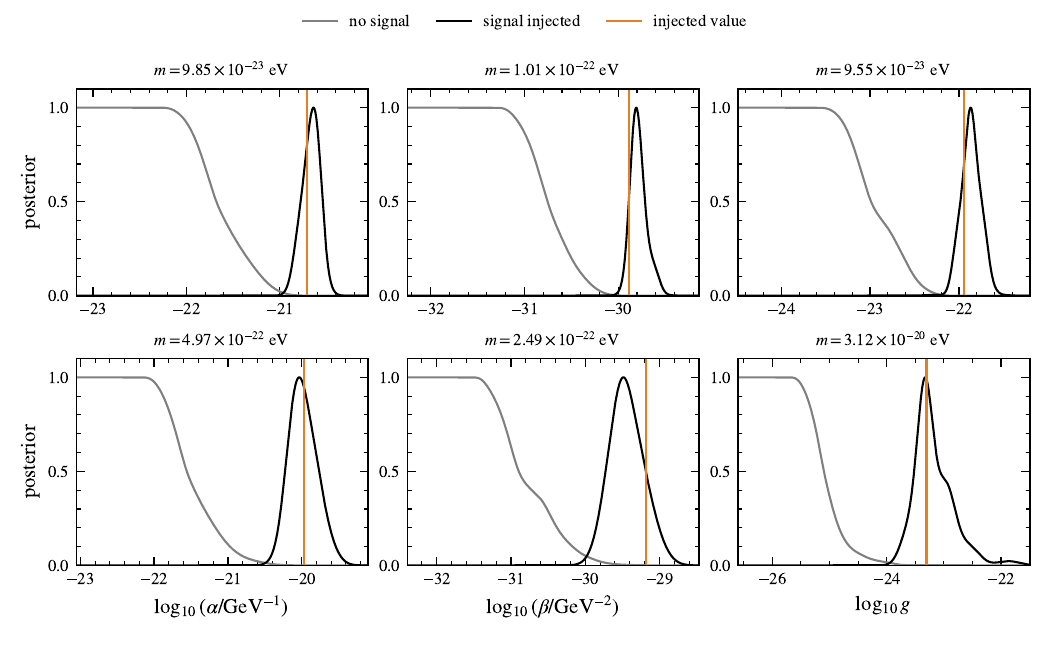}
    \caption{Combined posterior of the coupling for simulated data without a signal (grey) and with a signal injected at four times the limit on the real data (black), for the linear (left), quadratic (center) and vector (right) couplings, away from resonance (top) and close to a resonance (bottom). The vertical line marks the injected value. Both use a log-uniform prior on the coupling and each curve is normalized to its maximum.}
    \label{fig:injection}
\end{figure*}

\section{Orbital resonances and the treatment of \texorpdfstring{$\dot P_b$}{Pbdot}}
\label{app:pbdot}

The dips of Fig. \ref{fig:constraint-g} come from the resonant terms of the period channel, at $m\simeq\omega_b$ for the circular systems, and our mass grid resolves them down to their center. Here we describe the variations of the orbital parameters close to resonance, and show that holding $\dot P_b$ fixed or marginalizing it changes the limit only near the center of each resonance.

Let $\Delta\equiv |m-\omega_b|$. Keeping in Eq. \eqref{eq:vector_circular} only the terms in $m-\omega_b$ and expanding for $\Delta\,\bar t\ll1$,
\begin{subequations}\label{eq:vector_resonance}
\begin{align}
\frac{\overline{\delta P_b}}{P_b}
&\simeq\frac{3F_0}{4a\omega_b}\sin\vartheta\cos(\Upsilon_{\rm asc}+\varphi_{\rm asc})\,\bar t,\\
\delta T_{\rm asc}
&\simeq-\frac{F_0}{a\omega_b^2}\Bigl[\sin\vartheta\sin(\Upsilon_{\rm asc}+\varphi_{\rm asc})
\nonumber\\
&\qquad+\frac{\cot\iota\cos\vartheta}{2}\cos\Upsilon_{\rm asc}\Bigr]\bar t,\\
\frac{\delta x}{x}
&\simeq\frac{F_0}{a\omega_b}\Bigl[\sin\vartheta\cos(\Upsilon_{\rm asc}+\varphi_{\rm asc})
\nonumber\\
&\qquad-\frac{\cot\iota\cos\vartheta}{2}\sin\Upsilon_{\rm asc}\Bigr]\bar t,
\end{align}
\end{subequations}
up to corrections of order $\Delta\,\bar t$. Each of these drifts has the form of a timing parameter. A linear drift of $T_{\rm asc}$ is a constant shift of the period and is absorbed by the fitted $P_b$. A linear drift of the mean period has the time dependence of a constant $\dot P_b$, that is, of a term quadratic in time in the orbital phase; this is the term $\Xi_2$ of Refs. \cite{Kus2024, Huxhagen2026} (Appendix E of the former), which the forecasts there take as known and subtract. The drift of $x$ has the form of $\dot x$, which is absorbed in the systems that fit it. The same expansion applies, harmonic by harmonic, to the eccentric systems, and for the scalar couplings only the phase $\Upsilon$ enters the amplitude of the drift.

Equations \eqref{eq:vector_resonance} hold while $\Delta\,T_{\rm obs}\lesssim1$, the regime of constant drift assumed in Ref. \cite{Blas2019}. Once the detuning exceeds the frequency resolution of the data, $\Delta\,T_{\rm obs}\gtrsim2\pi$, the factors $\cos[(m-\omega_b)\bar t+\dots]$ of Eq. \eqref{eq:vector_circular} complete at least one oscillation within $T_{\rm obs}$. They can then no longer be written as a low-order polynomial in time, and the timing model cannot absorb them.

With $\dot P_b$ held at its published value, as in the rest of this work, the drift of the period survives the fit, with an amplitude proportional to $\sin\vartheta\cos(\Upsilon_{\rm asc}+\varphi_{\rm asc})$. At the center of the resonance the whole signal reduces to this single drift, whose size depends on one combination of the direction and phase of the field, and for a set of realizations this combination vanishes. Away from the center these factors run through all phases during the observations, and no realization cancels them. Since we marginalize over the direction and phase, the credible limit at the center is set by the realizations that switch the drift off, and the center of each dip is shallower than its deepest point. This loss of sensitivity comes from the unknown realization of the field and appears whatever the treatment of $\dot P_b$. It is the same effect that makes the vector resonances shallower than forecast (Sec. \ref{sec:results}).

If $\dot P_b$ is marginalized instead, it absorbs the drift of the period, and the limit also degrades near the center of the resonant peak. Reference \cite{Kus2024} found this behavior in a forecast for PSR J1903$+$0327 (their Appendix E). We test it here on the data, with the full timing model: we repeated the chains of the nine binaries that fit $\dot P_b$ with $\dot P_b$ marginalized together with the rest of the timing model, and recombined them with the chains of the other $41$ binaries.

Figure \ref{fig:pbdot} shows the result for PSR J1909$-$3744, whose resonance is the deepest of the nine. Away from the center the two limits coincide, and at the deepest point they agree to within $12\%$. Only near the center does the limit with a free $\dot P_b$ degrade, by more than an order of magnitude. Over the nine systems, the deepest point of each resonance moves by at most a factor of $5$. These shifts are comparable to the scatter between independent chains, and none of the comparisons of Sec. \ref{sec:results} changes. Holding $\dot P_b$ fixed therefore affects only a narrow band at the center of each resonance.
\section{Injection and recovery}
\label{app:injection}

As a check of the implementation, we inject a known signal into simulated data and analyze them with the same pipeline. For every binary, the residuals are replaced by a realization of the noise model of Sec. \ref{sec:noise}, to which we add the template of Eq. \eqref{eq:template-explicit} with a common coupling $\xi_{\rm inj}$. Each binary draws its own amplitude, phase and, for the vector coupling, direction of the field from their priors.

For each coupling we use two masses: one away from every orbital resonance, close to $10^{-22}\,$eV, and one close to a resonance of PSR J1903$+$0327 (scalar couplings) or PSR J1909$-$3744 (vector coupling). At each mass we analyze the same noise realization twice, without a signal and with $\xi_{\rm inj}=4\,\xi_{95}$, where $\xi_{95}$ is the limit of Sec. \ref{sec:results} at that mass. The coupling has a prior uniform in $\log_{10}\xi$ over the range of Table \ref{tab:priors}.

Figure \ref{fig:injection} shows the result. Without a signal the posterior is flat where no binary is sensitive and falls where the data exclude the coupling. With the signal the posterior peaks close to the injected value in all six cases.

\bibliographystyle{apsrev4-2}
\bibliography{referencias}

\end{document}